\documentclass[aps,prb,preprint,superscriptaddress]{revtex4-2}
\usepackage[T1]{fontenc}
\usepackage[utf8]{inputenc}

\usepackage[english]{babel}
\usepackage{graphicx}
\usepackage{hyperref}
\usepackage[section]{placeins}
\usepackage{float}\usepackage{multirow}%
\usepackage{amsmath,amssymb,amsfonts}%
\usepackage{amsthm}%
\usepackage{mathrsfs}%
\usepackage[title]{appendix}%
\usepackage{xcolor}%
\usepackage{textcomp}%
\usepackage{booktabs}%
\usepackage{algorithm}%
\usepackage{algorithmicx}%
\usepackage{siunitx} %
\usepackage{algpseudocode}%
\usepackage{listings}%
\newcommand{\beginsupplement}{%
  \setcounter{figure}{0}%
  \setcounter{table}{0}%
  \setcounter{equation}{0}%
  \setcounter{section}{0}%
  \renewcommand{\thefigure}{S\arabic{figure}}%
  \renewcommand{\thesection}{S\arabic{section}}%
  \renewcommand{\thetable}{S\arabic{table}}%
  \renewcommand{\theequation}{S\arabic{equation}}%
}
\theoremstyle{thmstyleone}%
\theoremstyle{thmstyletwo}%
\theoremstyle{thmstylethree}%
\begin{document}

\title{Localized emission in MoSe$_2$ monolayers on GaN nanopillars}

\author{Abderrahim Lamrani Alaoui}
\affiliation{Universit\'e C\^ote d'Azur, CNRS, CRHEA, Valbonne, France}

\author{\'Alvaro Moreno}
\affiliation{ICFO, Institute for Photonic Sciences, Castelldefels, Spain}

\author{Maximilian Heithoff}
\affiliation{ICFO, Institute for Photonic Sciences, Castelldefels, Spain}

\author{Virginie Br\"andli}
\affiliation{Universit\'e C\^ote d'Azur, CNRS, CRHEA, Valbonne, France}

\author{Aimeric Courville}
\affiliation{Universit\'e C\^ote d'Azur, CNRS, CRHEA, Valbonne, France}

\author{Maksym Gromovyi}
\affiliation{Universit\'e C\^ote d'Azur, CNRS, CRHEA, Valbonne, France}

\author{S\'ebastien Chenot}
\affiliation{Universit\'e C\^ote d'Azur, CNRS, CRHEA, Valbonne, France}

\author{Mahima-Ravi Srivastava}
\affiliation{Universit\'e C\^ote d'Azur, CNRS, CRHEA, Valbonne, France}

\author{St\'ephane V\'ezian}
\affiliation{Universit\'e C\^ote d'Azur, CNRS, CRHEA, Valbonne, France}

\author{Benjamin Damilano}
\affiliation{Universit\'e C\^ote d'Azur, CNRS, CRHEA, Valbonne, France}

\author{Frank Koppens}
\affiliation{ICFO, Institute for Photonic Sciences, Castelldefels, Spain}
\affiliation{ICREA, Instituci\'o Catalana de Recerca i Estudis Avan\c{c}ats, Barcelona, Spain}

\author{Yannick Chassagneux}
\affiliation{LPENS, ENS, CNRS, PSL, Sorbonne Universit\'e, Universit\'e Paris Cit\'e, Paris, France}

\author{Christophe Voisin}
\affiliation{LPENS, ENS, CNRS, PSL, Sorbonne Universit\'e, Universit\'e Paris Cit\'e, Paris, France}

\author{Philippe Boucaud}
\affiliation{Universit\'e C\^ote d'Azur, CNRS, CRHEA, Valbonne, France}

\author{Antoine Reserbat-Plantey}
\email{antoine.reserbat@cnrs.fr}
\affiliation{Universit\'e C\^ote d'Azur, CNRS, CRHEA, Valbonne, France}

\begin{abstract}
\textbf{
Solid-state quantum emitters (QEs) in two-dimensional semiconductors offer compact, chip-compatible sources for quantum photonics. In transition-metal dichalcogenides (TMDs), nanopillars are widely used to induce localized emission, yet the underlying confinement mechanism and the relative roles of strain versus dielectric environment remain unclear. The general problem addressed here is whether strain alone explains quantum emitter formation and placement in MoSe$_2$, or whether dielectric contrast at suspended-supported interfaces is also required. Here, we combine hyperspectral superlocalization of photoluminescence with co-registered AFM topography and phase to map the positions of localized states (LS) in MoSe$_2$ suspended on GaN pillars and correlate them with bending strain and the local dielectric context. Contrary to the common assumption of purely strain-driven activation, LS frequently occur at suspended--supported interfaces around the pillar apex and span a broad strain range without a clear threshold, while being scarce along high-strain ripples. Our data indicate that deterministic emitter positioning in Mo-based TMDs benefits from co-engineering both strain gradients and nanoscale dielectric heterogeneity, rather than strain alone. More broadly, this combined optical-mechanical characterization approach provides a general framework for mapping structure-property relationships in 2D quantum materials at the single-emitter level.}
\end{abstract}

\maketitle
\section{Introduction}\label{sec1}
 Solid-state quantum emitters~\cite{aharonovich_solid-state_2016} are central to integrated quantum photonics, enabling non-classical light sources—most notably single-photon sources that produce controllable photon streams with tailored quantum correlations. Different applications impose different requirements: platforms that demand highly indistinguishable photons include quantum computing, quantum simulation, and boson sampling, whereas indistinguishability is less critical for quantum-secure communication and certain metrology tasks. Multiple solid-state platforms have emerged, including fluorescent atomic defects and semiconductor quantum dots. Key challenges remain: scalable fabrication and integration, inhomogeneous and homogeneous broadening, and efficient photon extraction from hosts with a high refractive index.
 
Low-dimensional systems such as 2D materials~\cite{azzam_prospects_2021}, one-dimensional carbon nanotubes~\cite{he_carbon_2018} (CNTs), and zero-dimensional molecular emitters (e.g., nanographenes~\cite{zhao_single_2018}, molecules~\cite{toninelli_single_2021}) offer complementary advantages. They are resource-efficient (functional at the single-layer limit), compatible with low-footprint/low-energy operation, and accessible through both top-down and bottom-up fabrication routes. Several of these systems naturally extend the emission into the near-infrared/telecom bands (e.g., large-diameter CNTs and nanographenes). Critically, they can be back-end integrated onto photonic integrated circuits~\cite{akinwande_graphene_2019}.

Quantum emitters in semiconducting transition-metal dichalcogenides (TMDs) are particularly compelling~\cite{azzam_prospects_2021}. Initially observed in WSe$_2$ as spontaneously occurring emitters within both epitaxial~\cite{he_single_2015} and exfoliated flakes~\cite{tonndorf_single-photon_2015, chakraborty_voltage-controlled_2015, koperski_single_2015, srivastava_optically_2015}, they were soon engineered deterministically using AFM indentation/scratching~\cite{tonndorf_single-photon_2015, rosenberger_quantum_2019} and nanostructured substrates (nanopillars~\cite{palacios-berraquero_large-scale_2017, branny_deterministic_2017, kim_position_2019, luo_deterministic_2018}, pyramids~\cite{kim_position_2019}, slits~\cite{kern_nanoscale_2016}). A prevailing picture is that local strain softens the excitonic bands; in W-based TMDs, where the lowest exciton is dark, hybridization with defect states can break selection rules and activate bright recombination in both K and K$’$ valleys, an intervalley defect-exciton (IDE) process~\cite{linhart_localized_2019}. Additional routes include direct defect creation (e.g., ion~\cite{klein_site-selectively_2019}, ultra-violet light~\cite{wang_utilizing_2022}, or electron beams~\cite{parto_defect_2021, fournier_position-controlled_2021}), which has shown rapid progress in several hosts (notably thin hBN~\cite{fournier_position-controlled_2021}). A complementary and practical route is lateral heterostructuring: embedding MoSe$_2$ nano-islands within a WSe$_2$ monolayer creates lateral band-offset potential wells that act as quantum-dot–like emitters~\cite{kim_confinement_2024}. In parallel, van der Waals heterostructures have revealed excitons confined by moiré potentials~\cite{baek_highly_2020, seyler_signatures_2019}, opening a path to dense, ordered emitter arrays and collective optical phenomena~\cite{asenjo-garcia_exponential_2017} (e.g., selective radiance and subradiance) of interest for quantum memories and simulation.

\vspace{1em}

For quantum photonic architectures, deterministic control over emitter energy and position is crucial—especially for sub-$\lambda$ arrays that harness strong dipole–dipole interactions. This, in turn, requires both spectral uniformity (tuning emitters into mutual resonance) and spatial order (minimizing positional disorder). Conceptually, positioning QEs in 2D semiconductors reduces to engineering nanoscale excitonic confinement potentials. Electrostatic approaches can generate potentials varying over $\sim$10–50 nm and have yielded 1D~\cite{thureja_electrically_2022, heithoff_valley-hybridized_2024, hu_quantum_2024} and even 0D confinement~\cite{thureja_electrically_2024} in TMDs, though device complexity and cross-talk pose challenges. Alternative strategies include leveraging ferroelectric domain walls~\cite{soubelet_strong_2025}, sub-20-nm metagates~\cite{barcons_ruiz_engineering_2022}, and controlled patterning of the dielectric environment~\cite{raja_coulomb_2017, itzhak_exciton_2025}. The latter modifies both the quasiparticle gap and exciton binding energy via dielectric screening, producing red or blue shifts of excitonic transitions~\cite{ben_mhenni_breakdown_2025}; shaping the local environment thus provides a practical route to excitonic potential engineering~\cite{itzhak_exciton_2025}.

Strain engineering is likely the most common method for generating quantum emitters in TMDs. This is often achieved through the use of protrusions, such as nanopillars, and demonstrated arrays of emitters with a high degree of reproducibility. Pillars are typically made of SiO$_2$, Al$_2$O$_3$, PMMA, or gold when plasmonic enhancement is desired. Quantum emitters have been successfully demonstrated using such techniques in WSe$_2$, MoS$_2$, MoTe$_2$, and a few realizations in MoSe$_2$ using small indents~\cite{yu_site-controlled_2021}. Superlocalization $\mu$-PL methods, such as weighted spectral averaging, have been used to quantify emitter positions relative to the strain landscape by registering emission centroids with respect to nanopillars~\cite{branny_deterministic_2017} or photonic features such as waveguide edges~\cite{blauth_coupling_2018}. Although the pillar apex is often presumed to host the strain maximum, superlocalization of emitter positions combined with AFM-derived strain maps have revealed spatial displacements of hundreds of nanometers between emitters and the apex~\cite{xu_subdiffraction_2024}. Complementary analyses—including cathodoluminescence~\cite{luo_imaging_2023} and careful extraction of strain fields around indents~\cite{abramov_photoluminescence_2023}—support models in which defect states hybridize with strain-softened excitons (the intervalley defect–exciton picture), underscoring the need to co-engineer both the defect landscape and the strain profile for deterministic, high-quality QEs.

\vspace{1em}

Here, we directly correlate the nanoscale positions of localized states, extracted by superlocalization of hyperspectral $\mu$-PL, with the local strain field reconstructed from high-resolution AFM. We introduce a nonstandard sample architecture in which an exfoliated monolayer MoSe$_2$ is transferred onto high-quality GaN nanopillar arrays. This platform reproduces prior demonstrations of pillar-induced QE formation while enabling co-registered mapping of strain and dielectric contrast (via AFM phase). Our analysis suggests that strain alone may not account for the observed emitter statistics; correlations with suspended–supported interfaces are consistent with a contributory role of the dielectric environment alongside strain. Focusing on MoSe$_2$—less explored than WSe$_2$ for QE generation and featuring a bright lowest exciton at cryogenic temperature—offers a useful testbed to assess how strain gradients and local dielectric screening influence localization and brightness. Taken together, our observations are consistent with a combined strain–dielectric picture for deterministic positioning and spectral tuning of QEs in 2D semiconductors.

\section{Localized emission in MoSe$_2$ on GaN nanopillars.}\label{sec2}

\begin{figure}[H]
    \centering
    \includegraphics[width=1\linewidth]{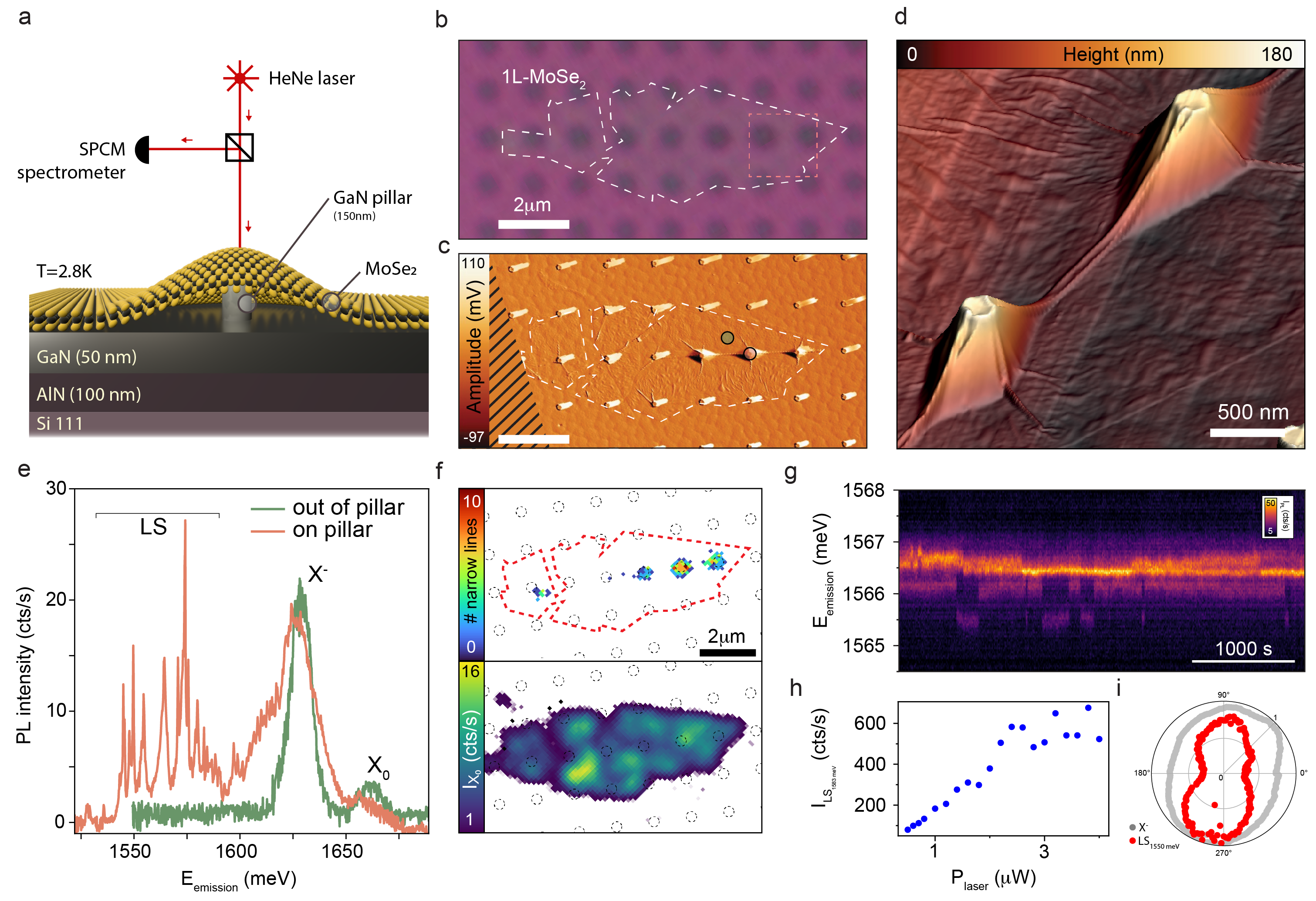}
    \caption{\textbf{Localized quantum emitters in monolayer MoSe$_2$ on AlN/GaN nanopillars.}  
        \textbf{a}: Schematic of the sample and $\mu$-PL setup. The AlN/GaN substrate is patterned to form nanopillars, onto which monolayer MoSe$_2$ is exfoliated. Micro-photoluminescence is performed at 2.8 K.  
        \textbf{b–d}: Optical micrograph (\textbf{b}) and atomic force micrographs (\textbf{c}, \textbf{d}) of the fabricated device.  
        \textbf{e}: Photoluminescence spectra recorded with the laser focused off (green) and on (orange) a nanopillar, with positions indicated in \textbf{c}. The spectra show neutral (X$_0$) and charged (X$^-$) excitons, along with a dense set of sub-meV localized states (LS) observed on the pillar.  
        \textbf{f}: Hyperspectral PL maps showing the spatial distribution of LS (top) and the X$_0$ intensity (bottom). LS appear exclusively at the pillar positions, highlighted by dashed circles.  
        \textbf{g}: Time-resolved PL spectrum of a single LS at 1566 meV, exhibiting spectral jumps of $\sim$500 $\mu$eV.  
        \textbf{h}: Power dependence of the PL intensity of a single LS, showing saturation behavior. 
        \textbf{i}: polarization-resolved PL revealing the linear polarization of LS emission (red), in contrast to the X$^-$ state (grey).}    
    \label{fig:fig1}
\end{figure}

We investigate localized emission in monolayer MoSe$_2$ transferred onto GaN nanopillar arrays using hyperspectral micro-photoluminescence ($\mu$-PL) at 2.8 K, correlated with high-resolution atomic force microscopy (AFM) (Fig.~\ref{fig:fig1}a). The GaN/AlN layers consist of 200 nm GaN grown by molecular beam epitaxy on a 100 nm AlN template atop a Si(111) substrate~\cite{semond_gan_1999}. Nanopillars are defined by e-beam lithography and dry etching, forming arrays of $\sim$150 nm-high GaN pillars with $\sim$160 nm diameter, and a remaining $\sim$50 nm GaN base layer.
We chose epitaxial GaN rather than the more common amorphous SiO$_2$ because epitaxially grown III-V substrates have been shown to suppress spectral diffusion and blinking in TMD quantum emitters compared to oxide substrates~\cite{iff_substrate_2017}, and to maximize dielectric contrast in regions where the TMD is locally suspended. Additionally, both GaN and AlN exhibit optical transparency in the relevant spectral range (excitation at 633~nm and emission at 750--800~nm), allowing efficient optical access. Details of sample and substrate preparation are provided in Supplementary Information note~\ref{si:sample_prep}.
Low-force AFM reveals that the MoSe$_2$ monolayer conforms to the nanopillars and forms tent-like suspended regions around them, together with ripples that connect neighboring pillars and radiate outward (Fig.~\ref{fig:fig1}b-c). These features reflect the competition between adhesion to the substrate and membrane bending. In our samples, the morphology indicates relatively strong adhesion: the membrane largely follows the GaN topography, with suspension occurring locally near the pillars and along ripples to minimize total energy.
We therefore selected the pillar pitch and aspect ratio to prevent full-span suspension across the array and to reduce the risk of membrane piercing associated with overly narrow pillars.
$\mu$-PL spectra acquired off-pillar display the characteristic MoSe$_2$ response (Fig.~\ref{fig:fig1}e): a neutral exciton (X$_0$) near 1.657 eV and a charged exciton (X$^-$) red-shifted by $\sim$50 meV, with positions and relative intensities varying across the flake (Fig.~\ref{fig:fig1}f, bottom). At pillar sites, in addition to X$_0$ and X$^-$, we observe multiple narrow, bright lines (sub-meV linewidth) red-shifted from the excitonic features (Fig.~\ref{fig:fig1}e). Following the literature, we refer to these as localized states (LS). These lines disappear above 20~K (cf. Supplementary Information note~\ref{si:Temp-dependence}), consistent with thermal detrapping of excitons for quantum emitters in TMDs~\cite{he_single_2015}.
\vspace{1em}

Hyperspectral mapping (one spectrum per laser position) enables automated counting of narrow lines across the field of view (Fig.~\ref{fig:fig1}f). The LS count is non-zero only at pillar locations and typically ranges from 3 to 9 per pillar under conservative detection thresholds (set to ensure robustness at modest signal-to-noise). With higher signal-to-noise ($>$10), we resolve up to 14 LS on a single pillar as discussed further in the Fig.~\ref{fig:fig4} for instance. These observations confirm that nanopillar arrays enable deterministic generation of localized emitters, as widely reported for WSe$_2$ and, more recently, for MoSe$_2$ (often via shallow indents).
While single-photon antibunching has been demonstrated in related architectures, it is more challenging in MoSe$_2$, likely due to short lifetimes. Our Hanbury Brown–Twiss setup has $\sim$300 ps timing resolution, which is insufficient for unambiguous continuous-wave $g^{(2)}(0)$ measurements; pulsed excitation upgrade would be required to resolve antibunching in this situation. Nevertheless, three signatures support the assignment of single-photon emitter-like behavior. First, we observe time-trace “spectral jumps” (Fig.~\ref{fig:fig1}g) with typical step sizes $\sim$0.6 meV, exceeding the emission linewidth ($\sim$150 $\mu$eV). These jumps are consistent with a localized zero-dimensional state coupled to a small number of nearby charge fluctuators (local Stark shifts). Extended excitonic emission, in contrast, drifts smoothly with power or field. Second, the localized lines also saturate with pump power—for example, reaching $\sim$600 cts/s at 4 $\mu$W under 633 nm excitation—consistent with a two-level system with a finite excited-state lifetime (Fig.~\ref{fig:fig1}h). Finally, their emission is strongly linearly polarized relative to the delocalized excitonic features such as X$^-$ states (Fig.~\ref{fig:fig1}i), indicating anisotropic confinement that lowers the in-plane symmetry, mixes valley states, and pins the dipole orientation. These signatures are not a substitute for an antibunching test, but together they strongly suggest the presence of individual quantum emitters that are both spatially confined at the nanoscale and spectrally isolated with sub-meV linewidths. We therefore refer to them as localized states (LS).

\vspace{1em}

At pillar sites, we consistently observe localized emission. A common interpretation invokes an IDE picture~\cite{linhart_localized_2019} in which local strain softens the excitonic bands and facilitates hybridization between defect states and excitons, breaking selection rules and activating emission. In W-based TMDs, where the lowest exciton is dark, this mechanism naturally accounts for bright, localized lines. In Mo-based compounds such as MoSe$_2$, the lowest exciton is already bright at cryogenic temperatures, so IDE is not strictly required to enable radiative recombination. The presence of intense, localized lines likely reflects additional ingredients - strain or electrostatic gradients, sharp variations in local dielectric screening — that funnel carriers into zero-dimensional states and enhance their radiative rate. 
In the following, we combine superlocalization of LS positions with AFM-derived strain maps to quantitatively assess how the local mechanical landscape—and, where relevant, dielectric contrast—correlates with emitter formation and spectral properties.

\FloatBarrier

\section{Extracting the bending-induced strain}\label{sec:strain}

\begin{figure}[H]
    \centering
    \includegraphics[width=1\linewidth]{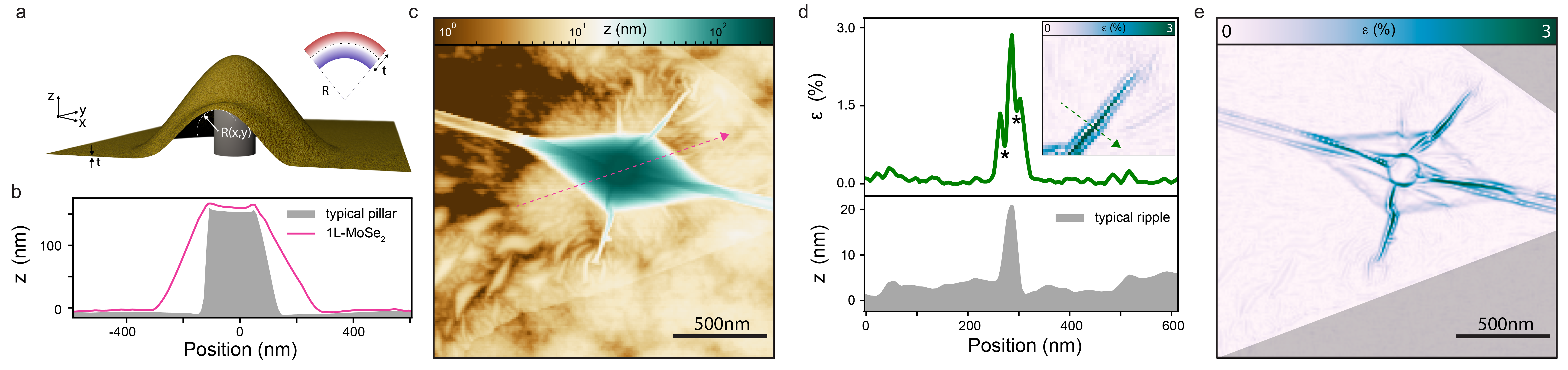}
\caption{\textbf{Bending-induced strain in monolayer MoSe$_2$ on a patterned GaN surface.}  
\textbf{a}: Schematic of a monolayer MoSe$_2$ flake (thickness $t$) transferred onto a GaN nanopillar. Depending on the balance between local adhesion energy and bending energy, MoSe$_2$ can conformally adhere to the substrate. Around the nanopillar, the monolayer becomes partially suspended, forming a tent-like shape. The local radius of curvature is denoted $R(x,y)$.  
\textbf{b}: Linecut of the MoSe$_2$ topography across a nanopillar (magenta), with position indicated in \textbf{c}. For comparison, a reference profile measured on a bare GaN nanopillar (grey, uncovered region) is overlaid without vertical shift.
\textbf{c}: Topography micrograph of the sample. The logarithmic scale enhances the visibility of transfer-induced features such as ripples, bubbles, and suspended regions around the pillar at the center.  
\textbf{d}: (Top) Extracted bending-induced strain $\epsilon$ across a ripple (see inset). (Bottom) Corresponding topography profile. The strain exhibits local minima (marked with *) corresponding to inflection points where the curvature changes sign—from convex (bending upward) to concave (bending downward).  
\textbf{e}: 2D map of the bending-induced strain in MoSe$_2$. The grey region corresponds to an area without MoSe$_2$, where strain extraction is not applicable.}
    \label{fig:fig2}
\end{figure}

We quantify the bending strain $\varepsilon(x,y)$ in a monolayer MoSe$_2$ membrane draped over GaN nanopillars by analyzing the AFM topography $z(x,y)$ (Fig.~\ref{fig:fig2}a). The membrane thickness ($t \sim 0.7\,\mathrm{nm}$) is much smaller than the local radii of curvature (typically $>\!30\,\mathrm{nm}$), allowing us to apply Kirchhoff--Love thin-plate theory under the small-slope approximation. We assume linear, isotropic elasticity and focus exclusively on bending, neglecting any in-plane pre-strain.

Low-force AFM reveals that MoSe$_2$ pins at pillar tops and becomes locally suspended around them, forming tent-like geometries with additional ripples that connect pillars and radiate outward (Fig.~\ref{fig:fig2}b,c). Over most suspended regions, the membrane exhibits cylindrical bending (one principal curvature dominates), allowing the strain to be expressed as
\begin{equation}
\label{eq:strain}
\varepsilon(x,y) \approx t |\omega| = \frac{t}{2}|\partial^{2}_x z + \partial^{2}_y z|\,,
\end{equation}
where $t \approx 0.7$~nm is the monolayer thickness and $\omega$ is the mean curvature. Details of the derivation and approximations are provided in the Supplementary Information note ~\ref{si:Bending-strain}.

We first validate the strain extraction on a single one-dimensional ripple (Fig.~\ref{fig:fig2}d). The topography shows a height of $\sim 20\,\mathrm{nm}$. The bending-strain magnitude $\varepsilon$ is near zero on both sides of the ripple, increases where the membrane bends upward, and remains positive when the curvature later inverts. At the inflection point, where bending changes from up to down, $\varepsilon$ should pass through zero. Experimentally, we observe a pronounced minimum (marked by * in Fig.~\ref{fig:fig2}d) that does not reach exactly zero, consistent with finite pixel size and line-cut averaging. The inset of Fig.~\ref{fig:fig2}d more clearly resolves the expected zero-strain lines along the ripple flanks where the curvature changes sign. This behavior matches the bending-strain profile expected across a 1D ripple.
Applying the same analysis around a pillar shows that $\varepsilon$ is largest along the circumference near the pillar apex and along ripples that radiate away from it. The bending-strain magnitude reaches up to $\sim 3\%$, consistent with values reported for similar nanopillar architectures~\cite{abramov_photoluminescence_2023}. Note that strain inferred from exciton energy shifts may differ from AFM-derived values because optical spectra are averaged over a $\sim 1~\mu\mathrm{m}$ excitation/collection spot. Independent AFM estimates of monolayer strain in related systems typically lie in the $0\text{--}4\%$ range~\cite{abramov_photoluminescence_2023}. Moreover, recent \emph{in situ} strain-tuning experiments place the onset of defect--exciton hybridization in the $1\text{--}3\%$ window~\cite{hernandez_lopez_strain_2022}, consistent with theoretical expectations for the IDE mechanism~\cite{linhart_localized_2019}. Elevated strain is also visible along the clamping line where the MoSe$_2$ membrane detaches from the substrate to form the tent-like suspended region. On the flat substrate, $\varepsilon$ is smaller but nonzero, correlating with short ripples and bubbles characteristic of dry transfer.
With the bending-strain field established across the sample, we now correlate it with the positions of the localized states (LS) obtained via PL superlocalization.

\FloatBarrier

\section{Super-localization of emitters}

Far-field optical microscopy is fundamentally limited by the diffraction barrier. Nevertheless, super-resolution techniques (e.g. STED, PALM, STORM) and super-localization approaches have made it possible to access mesoscale information down to 20–200 nm, depending on the method~\cite{schermelleh_super-resolution_2019}.
The super-localization technique relies on  deconvoluting the image with the point-spread function (PSF) to extract the centroid position of individual emitters with precision  below the diffraction limit. The precision of the extracted coordinates depends on the quality of the Gaussian fit to the emission spot, which ideally corresponds to a diffraction-limited Airy disk, and is therefore strongly linked to the signal-to-noise ratio.
This super-localization technique has been applied to emitters in carbon nanotubes~\cite{raynaud_superlocalization_2019} and in WSe$_2$~\cite{abramov_photoluminescence_2023} on nanoparticles~\cite{xu_subdiffraction_2024} and pillars~\cite{branny_deterministic_2017}, with position accuracies ranging from 15 to 120~nm, depending on the experimental signal-to-noise ratio and sampling conditions.
To apply this concept to our measurements, we acquire hyperspectral maps: the laser spot is scanned across the sample, and at each position, a PL spectrum is recorded. The maps are deliberately oversampled with a pixel size of $\sim$50~nm to improve the robustness of the Gaussian fits, and each hyperspectral acquisition can last over 48~h, requiring exceptional setup stability; during this time the emitters remain photostable, with well-separated spectral lines that prevent jitter-induced mixing.

\vspace{1em}

\begin{figure}[H]
    \centering
    \includegraphics[width=1\linewidth]{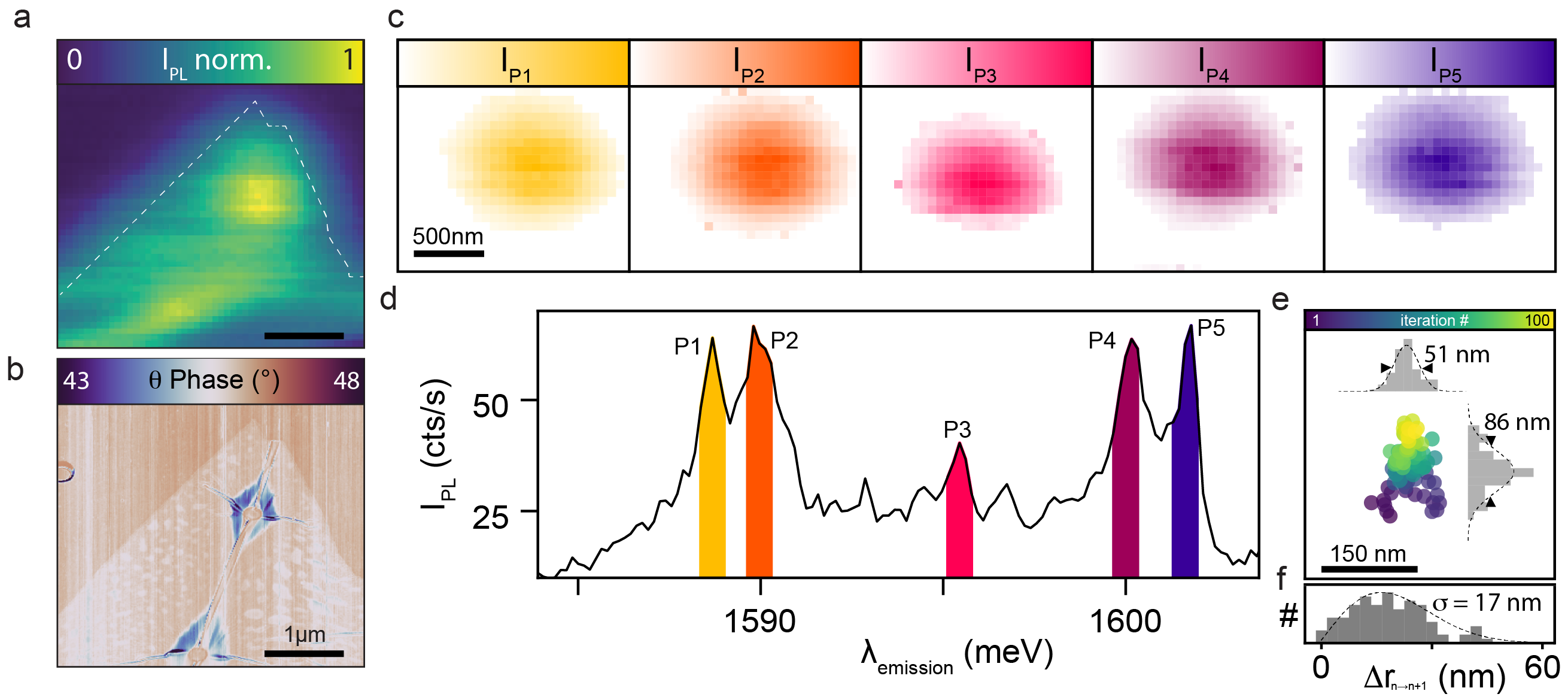}
\caption{\textbf{Super-localization of emitters in MoSe$_2$.}
\textbf{a}: Hyperspectral photoluminescence (PL) map integrated over the full spectral range; one spectrum is acquired at each scan position.
\textbf{b}: Atomic force microscopy (AFM) phase micrograph of the same region.
\textbf{c}: PL excitation map with spectral filtering of the emission according to the color code in (d); spot centroids are obtained from 2D spatial Gaussian fits.
\textbf{d}: Representative PL spectrum near a nanopillar position showing multiple localized states.
\textbf{e}: Super-localization map from 100 repeated scans with fitted emission centroids. Projected 1D histograms (top/right) show Gaussian-fitted spreads of \(\sigma_x \approx 51~\mathrm{nm}\) and \(\sigma_y \approx 86~\mathrm{nm}\) (laser spot: ~600 nm).
\textbf{f}: Histogram of the frame-to-frame displacement \(\Delta r_n = \lVert \mathbf{r}_{n+1}-\mathbf{r}_n\rVert\); the distribution peaks near \(17~\mathrm{nm}\), consistent with the single-step increment of the translation stage.
}
    \label{fig:fig3}
\end{figure}

An example of the total emission intensity over such a map is shown in Fig.~\ref{fig:fig3}a, alongside the AFM phase signal of the same region (Fig.~\ref{fig:fig3}b). A representative spectrum is displayed in Fig.~\ref{fig:fig3}d, from which we isolate several peaks (P1–P5) and extract their corresponding intensity profiles (Fig.~\ref{fig:fig3}c). As expected for localized emitters in a confocal microscope, these profiles exhibit Gaussian shapes. The centroid coordinates are then extracted from the fiiting procedure, and the results are discussed in Fig.~\ref{fig:fig4}.
To evaluate the stability and robustness of the method, we repeated the mapping 100 times on a bright scatterer, allowing faster acquisitions. For each map we extracted the emission centroid and overlaid all positions (Fig.~\ref{fig:fig3}e). The per-axis distributions yield standard deviations of \(\sigma_x \approx 51~\mathrm{nm}\) and \(\sigma_y \approx 86~\mathrm{nm}\), well below the diffraction limit. The iteration color code reveals a slow drift predominantly along \(y\), which inflates \(\sigma_y\).
To isolate mechanical jitter, we computed the frame-to-frame displacement \(\Delta r_n = \lVert \mathbf{r}_{n+1} - \mathbf{r}_n \rVert\). The corresponding histogram peaks at \(\sim 17~\mathrm{nm}\) (Fig.~\ref{fig:fig3}f), indicating that the typical inter-scan shift is on the order of a single step of the translation stage \(\sim 16~\mathrm{nm}\). Thus, the measured inter-frame motion approaches the single-step limit of our scanner. We also carefully assessed signal-to-noise ratio and blinking effects through numerical simulations (cf. Supplementary Information note~\ref{si:superloc-precision}), confirming that our high signal-to-noise and low fluctuation timescales ensure minimal impact on localization accuracy. 
A remaining challenge is the precise registration of these super-localized positions with the AFM micrograph. To achieve this, we performed larger-scale PL and reflectance imaging of the sample and aligned the central position of the pillar grid, from which we derived the global transformation matrix (including rotation, translation, and isotropic scaling). The full procedure is provided in the Supplementary Information note ~\ref{si:coordinate-system}.

\FloatBarrier

\section{Mechanisms for Localized Emitter Formation in MoSe$_2$}

\begin{figure}[H]
    \centering
    \includegraphics[width=1\linewidth]{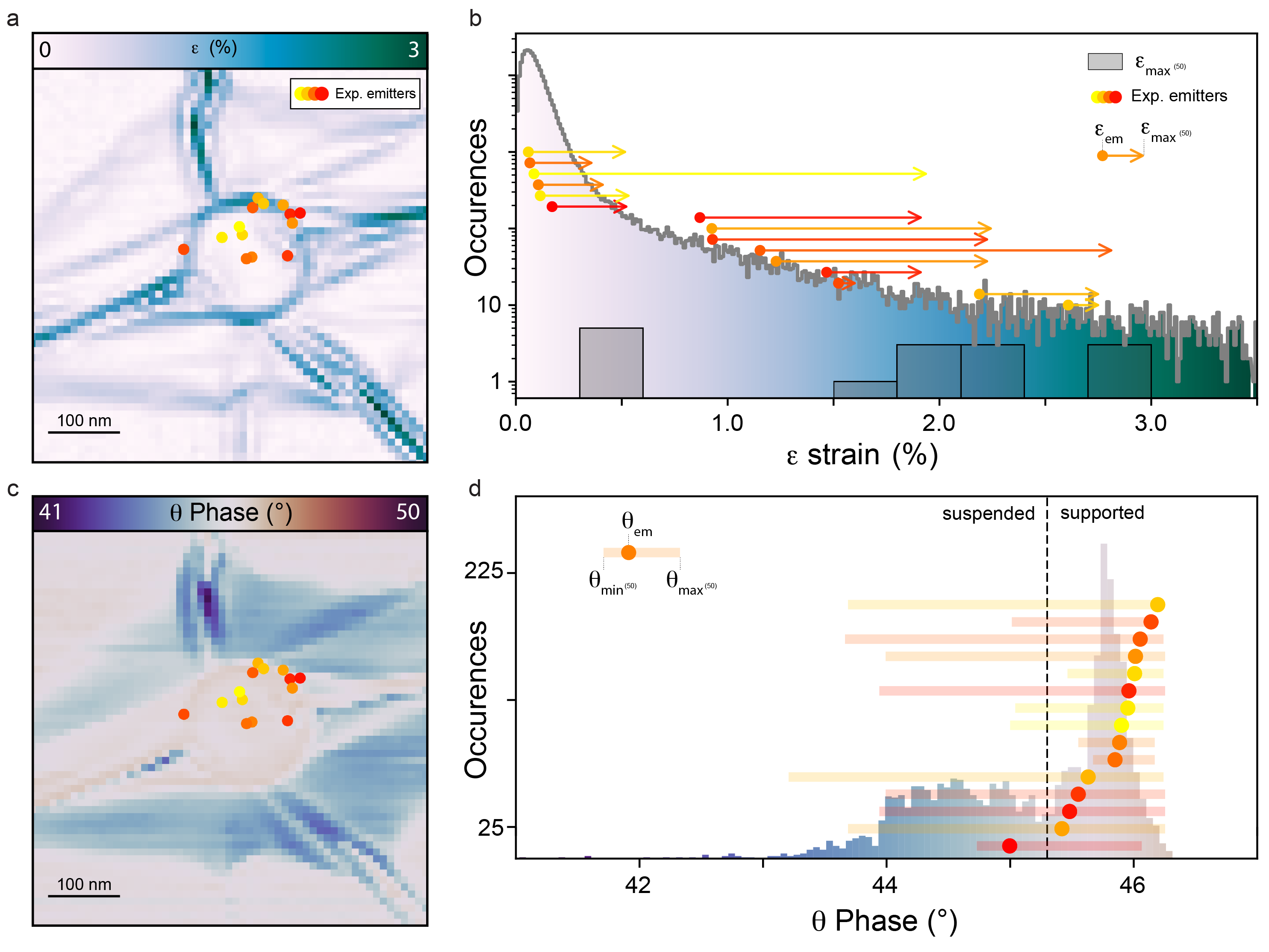}
\caption{\textbf{Impact of strain and dielectric environment on the positions of localized emitters in monolayer MoSe$_2$.}
\textbf{a}: Bending-induced strain map overlaid with the extracted positions of localized states (LS). Dot color encodes emission energy (yellow–red: 1540–1610 meV).
\textbf{b}: Histogram of the bending-induced strain. Colored markers indicate the strain at LS centroids; arrows show the local maximum strain within a 50 nm neighborhood of each centroid ($\epsilon_{\mathrm{max}}^{(50)}$), highlighting that emitters may experience higher strain than at the centroid itself. The distribution of $\epsilon_{\mathrm{max}}^{(50)}$ (grey) shows LS occurring across the full strain range.
\textbf{c}: AFM phase ($\theta$) image of the same region with LS positions overlaid. Blue regions correspond to suspended MoSe$_2$ (tent-like geometry from pillar top to substrate), whereas warmer tones indicate supported areas; the dielectric environment varies sharply across this interface.
\textbf{d}: Histogram of AFM phase $\theta$. The vertical dashed line marks the suspended–supported boundary. Colored points denote the phase at each LS centroid (same color code as in \textbf{a}); horizontal bars show the phase range within a 50 nm neighborhood. Most bars cross the boundary, indicating that LS frequently lie near suspended–supported interfaces, consistent with dielectric-contrast–assisted localization.}
    \label{fig:fig4}
\end{figure}

When comparing the super-localized emitter positions with the AFM map, we find that they consistently appear near the pillar apex, most often along the crown-shaped circumference. This crown-like feature likely originates from slight material extrusion during the masking step of the etching process, as visible in Fig.~\ref{fig:fig2}. Importantly, the emitters are not located along the ripples or at the base of the tent-like suspended regions (Fig.~\ref{fig:fig4}a). Within the pillar region itself, we observe no systematic correlation between the precise emitter position and its emission energy, in agreement with Ref.~\cite{xu_subdiffraction_2024}.

One possible concern is whether exciton diffusion could shift the apparent emitter location in our confocal measurements. Since excitation and detection occur at the same spot, excitons generated at a ripple might migrate before recombining radiatively elsewhere. However, this scenario is unlikely given the limited diffusion lengths under our experimental conditions. Exciton transport along 1D ripples in hBN-encapsulated WSe$_2$ has been reported to be anisotropic~\cite{Dirnberger2021}, with diffusion lengths on the order of 1~$\mu$m at 4~K after 1~ns. While a small fraction of such diffused emission could, in principle, be collected even if recombination occurs away from the excitation spot, the negligible contribution of strain funneling along ripples in our geometry makes significant exciton migration unlikely. The absence of LS along high-strain ripples therefore reflects their actual spatial distribution rather than a measurement artifact.

The lack of LS formation along ripples, despite hosting the largest bending strain, suggests that strain magnitude alone does not determine localization. One possibility is that LS require exciton energies to be tuned into resonance with defect states~\cite{hernandez_lopez_strain_2022, linhart_localized_2019}, which may only activate above a finite strain threshold (reported between 1\% and 2.5\% in WSe$_2$~\cite{hernandez_lopez_strain_2022}). Nonetheless, this interpretation is less straightforward for Mo-based TMDs, where hybridization is expected to involve bright rather than dark states, in contrast to W-based systems~\cite{linhart_localized_2019}. As a result, the exciton population should still predominantly decay through the fast bright-state channel, with only a small fraction recombining via defect-mediated interlayer excitons.

\vspace{1em}

To account for the spatial distribution of the localized states, we consider several limiting scenarios. 
In the first null scenario, emitters form via stochastic processes that are unrelated to local strain or structural features. In this case, their spatial distribution would be indistinguishable from a uniform random placement within the mapped area. This hypothesis can be ruled out, as we clearly observe a clustering of LS around the pillar apex.
To quantify this deviation from randomness, we simulated a uniform random distribution of emitters over the area shown in Fig.~\ref{fig:fig4}a and analyzed their occurrence as a function of the local bending-strain percentile. Experimentally, 9 emitters are found within the top 90th strain percentile (corresponding to $\varepsilon > 1.47\,\%$). A purely random distribution would yield on average $\sim$5 emitters in this region, corresponding to an enrichment factor of $\sim$1.8. The probability of observing 9 or more emitters in the top 90th percentile under a random distribution is $\sim 3 \, \%$, indicating that such clustering is unlikely to arise by chance alone. This statistical analysis, detailed in the Supplementary Information note ~\ref{si:stat_analysis}, demonstrates that LS formation is nonrandom and correlated with the local environment, with bending strain emerging as a primary contributor at this stage.
A second scenario assumes that emitters preferentially form in regions exceeding a critical strain threshold, consistent with the idea that strain softens the excitonic bands sufficiently to enable hybridization with defect states. However, the spatial distribution of LS contradicts this picture: even when accounting for a $\sim$50~nm positional uncertainty, the highest-strain regions—namely the one-dimensional ripples—do not host emitters.
To further evaluate the strain landscape experienced by the LS, we extracted the bending strain at each emitter position. To account for localization uncertainty, we also report the maximum strain within a 50 nm neighborhood around each emitter. These values are shown in Fig.~\ref{fig:fig4}b, together with the strain histogram of the monolayer region displayed in Fig.~\ref{fig:fig4}a. LS are observed across a broad strain range, from 0.1\% up to 2.8\%. Their distribution does not reveal distinct groupings, and a significant fraction of emitters is observed across low, intermediate, and high strain values. Importantly, the data reveal neither a minimum strain threshold nor a preferred strain value for LS formation, indicating that strain alone is insufficient to determine localization.

\vspace{1em}

The AFM measurements also provide access to the phase signal $\theta$, shown in Fig.~\ref{fig:fig4}c. The phase delay of the oscillating AFM cantilever in tapping-mode scanning corresponds to the nanoscale local energy dissipation. Consequently, AFM phase contrast reflects the mechanical environment of the 2D material~\cite{vasic_phase_2017}: membranes on hard versus soft substrates generally exhibit different phase values, and a clear variation is typically observed between supported and suspended regions. In our measurements, we find that the phase differs between the supported regions of the MoSe$_2$ monolayer at the top and bottom of the pillar and the suspended tent-like areas. Interestingly, along the ripple, the AFM phase indicates an increased apparent stiffness, comparable to that in the supported regions. This effect is consistent with tension-induced (geometric) stiffening of the suspended membrane.  

To quantify the phase contrast, we plot the histogram of $\theta$ on Fig.~\ref{fig:fig4}d, which reveals two regimes corresponding to supported regions ($\theta > 45.3^\circ$) and suspended ones ($\theta < 45.3^\circ$). While this threshold may vary over larger scan areas due to long-range mechanical inhomogeneities, in the reduced area studied here, the phase mapping provides a robust criterion for distinguishing suspended from supported zones.

\vspace{1em}

We then apply the same procedure as for the bending strain map: for each emitter, we extract the AFM phase value at its position, $\theta_{\mathrm{em}}$, and to better represent the positional accuracy we also report the span of phase values, $\theta_{\max^{(50)}} - \theta_{\min^{(50)}}$, within 50~nm around the emitter. This provides information on the local mechanical environment of the localized state. For example, when an emitter lies in a fully suspended or fully supported region, the phase values remain narrowly distributed around a central value. In contrast, when the emitter is located near an interface—such as a clamped edge where the membrane transitions from suspended to supported—the phase values extend over a broader range. No clear correlation is found between the emitter energy and the absolute phase value. However, we observe that $\sim$80\% of the 15 emitters are located at suspended/supported interfaces. This constitutes one of the key findings of this work: localized states preferentially appear at interfaces.  

Although we have shown no direct correlation with bending strain magnitude, the presence of an interface implies a sharp dielectric contrast. Indeed, GaN has a relatively high dielectric constant~\cite{levinstejn_properties_2001} ($\varepsilon_{\mathrm{GaN}} \sim 9$), so MoSe$_2$ excitons experience an abrupt change in dielectric environment when transitioning from vacuum (above and below the membrane) to GaN underneath. The relevant length scale for the sensitivity of TMD excitons to such dielectric variations is defined by the extent of their wavefunction, approximately given by the Bohr radius $r_B$. It is established, both theoretically and experimentally, that dielectric screening in low-dimensional systems~\cite{raja_coulomb_2017, chernikov_exciton_2014, xu_creation_2021} (e.g., 2D materials, nanotubes) strongly influences exciton transition energy. The precise microscopic mechanism remains under debate, as both the static dielectric component and high-frequency contributions~\cite{van_tuan_effects_2024, ben_mhenni_breakdown_2025} may play a role, affecting exciton binding energy and bandgap renormalization. 

\vspace{1em}

Similar questions regarding the relevant dielectric function have long been debated in the context of carbon nanotubes. Experimental studies systematically report a redshift of excitonic resonances with increasing dielectric constant of the environment~\cite{campo_optical_2021, walsh_scaling_2008}, although quantitative agreement with theory remains limited. One central difficulty is that the nanotube's inner and outer dielectric environments often differ, making it challenging to define an effective screening parameter. Theoretical analyses~\cite{ando_environment_2010} highlight that the choice between static and optical dielectric constants is nontrivial, with different arguments supporting each. These issues directly resonate with our case of MoSe$_2$ on GaN pillars, where excitons also probe dielectric variations at the nanometer scale, raising the same fundamental question of how to correctly describe dielectric screening as the characteristic exciton size approaches the spatial inhomogeneity of the environment.

In our case, if exciton confinement were to arise from dielectric effects, one would require a geometry capable of producing a point-like potential well. A simple one-dimensional interface, even when curved, between suspended and supported regions cannot by itself induce such confinement. Instead, a nanoscale dielectric inhomogeneity with localized contrast would be necessary. One possible candidate is the crown-like roughness at the pillar apex, revealed by AFM in Fig.~\ref{fig:fig2}b, which could in principle provide such confinement. However, without nanoscale investigations, such as EL-STM or visible cryo-SNOM, it is difficult to directly demonstrate the existence of such geometries. 

\vspace{1em}

The absence of localization at the bottom interface of the tent-like suspended region, where dielectric contrast is also present, may support this interpretation: near the apex, the monolayer is likely suspended over a granular or uneven surface, whereas at the bottom it rests on a continuous, flat GaN interface, producing a smoother 2D suspended/supported boundary. This asymmetry suggests that the microscopic mechanism is governed by a subtle interplay of local curvature, defect distribution, and dielectric screening. A central challenge, therefore, is to disentangle the relative contributions of bending strain and dielectric contrast. Since both mechanisms can, in principle, yield confinement potentials of comparable magnitude, a definitive understanding will require a combined analysis integrating independent strain mapping with controlled dielectric engineering.

One may also consider the reverse situation, namely, the absence of dielectric contrast. In fully encapsulated TMD monolayers, it is natural to ask whether quantum emitters can still emerge on top of pillars. In previous experiments on hBN-encapsulated WSe$_2$ placed on pillar arrays~\cite{parto_defect_2021}, no quantum emitters were observed unless the material was irradiated at the pillar positions using an electron beam. This result suggests either that the dielectric environment plays an essential role in the formation of localized states or that strain in the TMD layer is relaxed by a thicker stack. This possibility was explicitly investigated in another study~\cite{daveau_spectral_2020}, where WSe$_2$ monolayers with no or partial hBN encapsulation were transferred onto pillars. Quantum emitters were predominantly observed when only the top face was covered with hBN, leaving the bottom interface directly exposed to dielectric contrast. However, in two cases where a thin bottom hBN spacer was present, emitters were still detected. Notably, in that configuration, the bottom hBN consisted of a bilayer ($\sim$0.6~nm), thin enough that excitons in the TMD still experienced dielectric screening from the substrate.  

Further evidence comes from local indentation experiments, in which an AFM tip was used to strain hBN-encapsulated MoSe$_2$~\cite{gelly_probing_2022}. While strong redshifts of excitonic lines were observed, these shifts corresponded to free excitons drifting in a non-uniform strain field rather than to the formation of long-lived localized emitters. Overall, most experiments on quantum emitters in TMDs employ a geometry with pronounced dielectric contrast. For instance, WSe$_2$ deposited on silver nano-islands~\cite{tripathi_spontaneous_2018}, on pillars~\cite{palacios-berraquero_large-scale_2017, branny_deterministic_2017}, or in pits fabricated by gallium FIB ion milling~\cite{yu_site-controlled_2021}, which are prone to residual contamination at the bottom. It is worth noting that other classes of localized excitons—such as those confined by moiré potentials~\cite{baek_highly_2020, seyler_signatures_2019} or created by defect implantation~\cite{klein_site-selectively_2019}—arise from distinct mechanisms, either through periodic potential modulation or direct defect engineering, and should be considered separately from the strain–dielectric scenarios discussed here.

\vspace{1em}

Patterned substrates with high dielectric contrast, achieved through nanoscale holes~\cite{barcons_ruiz_engineering_2022, itzhak_exciton_2025} or deposited nanoparticles, may enable engineered confinement sites for excitons in TMD monolayers. The dielectric environment modulates exciton energies over length scales comparable to the exciton wavefunction extension (~1–2 nm for the 1s exciton in MoSe$_2$), as demonstrated by the 30 meV blueshift observed for WSe$_2$ on SrTiO$_3$ compared to hBN substrates~\cite{ben_mhenni_breakdown_2025}. Abrupt dielectric boundaries can create sharp confinement potentials analogous to those in core-shell quantum dots, in which interface engineering influences Auger recombination and multi-exciton dynamics~\cite{park_effect_2014}. Such dielectric engineering could offer advantages over defect-mediated localization by enabling deterministic, large-scale fabrication with subwavelength periodicity, which is relevant for future exploration of collective phenomena in quantum emitter arrays~\cite{asenjo-garcia_exponential_2017}.

\FloatBarrier

\section{Conclusion}\label{sec13}

We combined hyperspectral superlocalization of $\mu$-PL with co-registered AFM topography and phase and found that most localized states arise at suspended–supported interfaces around the pillar apex, spanning a broad range of bending strains, with no clear threshold.  This discrepancy, together with the interface clustering, suggests a cooperative mechanism in which nanoscale dielectric heterogeneity directs strain-softened excitons into zero-dimensional states. These insights motivate a co-design strategy for deterministic emitters in Mo-based TMDs that jointly engineer strain gradients and dielectric contrast—via pillar geometry, apex roughness, and controlled spacer layers—rather than relying solely on strain. Looking ahead, correlating our maps with nanoscale probes of defect states (e.g., EL-STM or cryogenic s-SNOM), implementing tunable screening (thickness-graded hBN or metagates), and performing $g^{(2)}(0)$ antibunching measurements under pulsed excitation on single lines will clarify the respective roles of defects, curvature, and screening in setting brightness, lifetime, and indistinguishability. Ultimately, this methodology, which combines superlocalization, strain mapping, and dielectric contrast analysis, provides a systematic approach to engineer deterministic sub-$\lambda$ emitter arrays with controlled spectral properties, thereby advancing the scalability of integrated quantum photonic platforms.

\section*{Supplementary information}
\beginsupplement
\section{Sample preparation and AFM measurements}
\label{si:sample_prep}
The GaN (200 nm) / AlN (100 nm) sample was grown by molecular beam epitaxy on a Si(111) substrate in a RIBER reactor, similar to the approach described in Ref.~\cite{le_louarn_aln_2009}. The group-III elements are supplied by solid-source effusion cells, and the nitrogen precursor is ammonia. The growth temperatures are 780~$^\circ$C for GaN and 900~$^\circ$C for AlN.
Atomic force microscopy (AFM) measurements of monolayer MoSe$_2$ on AlN/GaN nanopillar arrays were performed with a Bruker Dimension Edge operated in tapping mode. To prevent damage to the monolayer while resolving the complex nanostructure, scans were acquired at very slow speeds (0.1~Hz line frequency, corresponding to a 90-minute image) with carefully optimized intermittent tip--sample contact (setpoint = 4.7~V).
Arrays of GaN pillars were fabricated by reactive ion etching (RIE) using a nickel (Ni) hard mask formed via a lift-off process based on a positive PMMA e-beam resist. Prior to lithography, the samples were sonicated  in acetone and isopropanol (IPA), followed by dehydration baking on a hot plate at 125~$^\circ$C. A 100~nm-thick layer of PMMA 495~K A4 resist was spin-coated using an RC8 Gerset coater. Electron-beam lithography was carried out at 20~kV using a Raith ELPHY PLUS system integrated with a Zeiss Supra~40 scanning electron microscope. To ensure smooth and well-defined feature edges, a step size of 5~nm was employed in combination with the Raith circle-writing mode. The exposure dose and development time were optimized to achieve a resist profile suitable for clean lift-off. Development was performed in a 1:3 MIBK/IPA solution.
A 30~nm-thick Ni layer was deposited using an EVA450 e-beam evaporator, followed by lift-off in acetone. The detached Ni flakes were removed by gentle pipette rinsing in acetone, avoiding ultrasonic agitation. The sample was subsequently rinsed in IPA and subjected to UV-ozone cleaning to eliminate resist residues. The resulting 30~nm Ni mask was used to etch 150~nm-high GaN pillars in a chlorine-based plasma using an Oxford System~100 RIE-ECR reactor. After etching, the Ni mask was removed in a piranha solution (H$_2$SO$_4$/H$_2$O$_2$ = 3:1), which also served as a final wet cleaning step prior to the subsequent 2D material transfer.

\section{Temperature-dependent photoluminescence of Localized States}
\label{si:Temp-dependence}
\begin{figure}[h]
    \centering
    \includegraphics[width=0.95\linewidth]{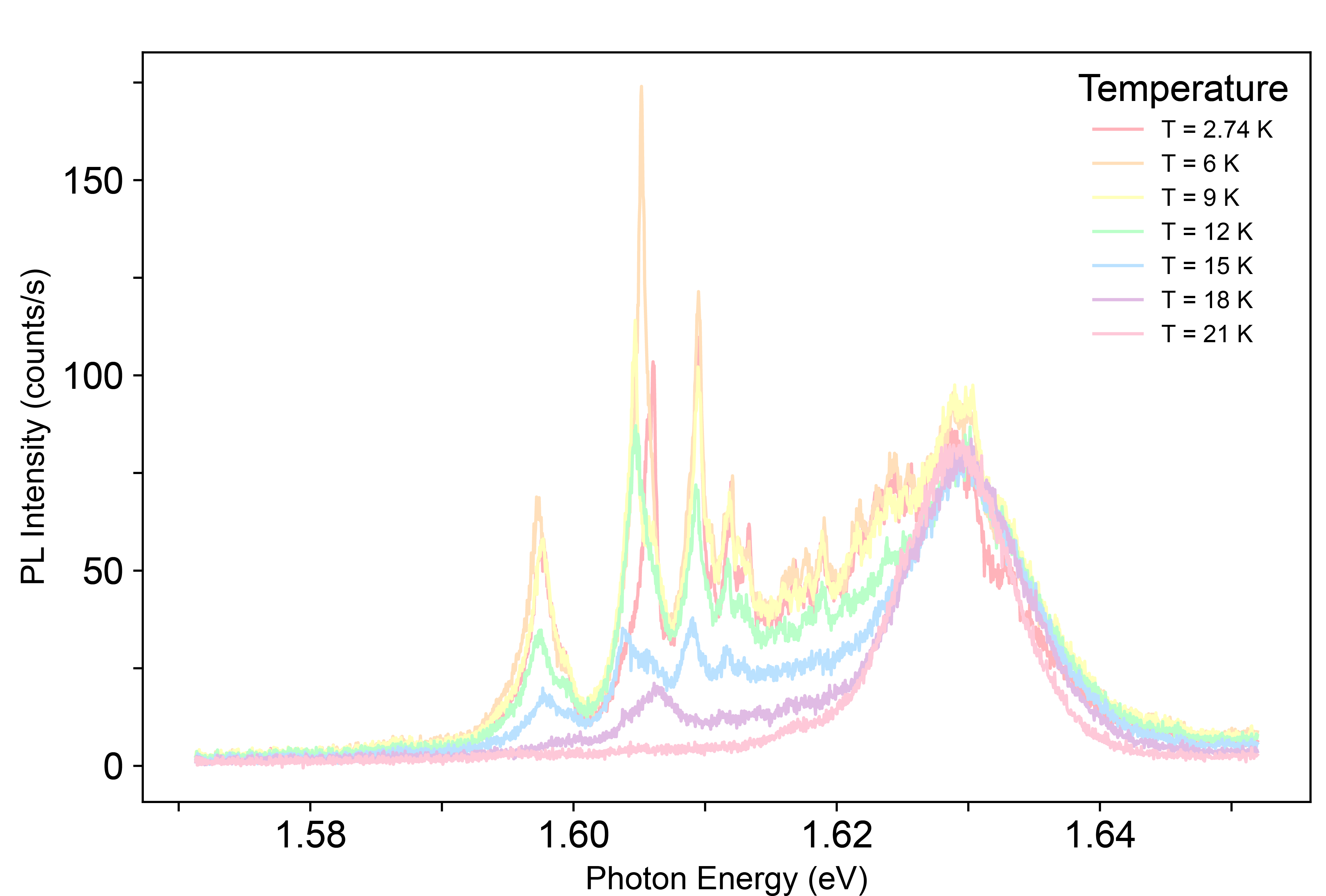}
    \caption{Temperature-dependent photoluminescence spectra of MoSe$_2$ on top of a pillar. (a) PL spectra showing narrow emission lines red-shifted from the free-exciton energy, associated with localized states. The spectra are shown for temperatures ranging from 2.74 K to 21 K. Above 20 K, no localized state emission is observed. At each temperature, emission maps were acquired to reposition the laser spot at the same location and compensate for thermal drift of the sample holder (although drift is minimal in the 0-30 K range for this specific setup).}
    \label{fig:T-dependent-spectra}
\end{figure}

To investigate the thermal stability of the observed localized states, we performed temperature-dependent photoluminescence measurements in the range 2.74-21 K. Figure~\ref{fig:T-dependent-spectra} shows the evolution of the PL spectra as a function of temperature. At the lowest temperature (2.74 K), we observe several narrow emission lines in the energy range 1.59-1.62 eV, red-shifted from the free-exciton emission, characteristic of localized states. As temperature increases, these lines progressively decrease in intensity and eventually disappear above 20 K, consistent with thermal detrapping of excitons from the localization sites, as reported in most studies on quantum emitters in transition metal dichalcogenides.

\FloatBarrier

\section{Derivation of bending-induced strain from AFM topography}
\label{si:Bending-strain}
This section provides the detailed derivation of the bending strain expression used in the main text (Eq.~\ref{eq:strain}). We extract the bending-induced strain field $\varepsilon(x,y)$ from the measured AFM topography $z(x,y)$ of monolayer MoSe$_2$ draped over GaN nanopillars using Kirchhoff--Love thin-plate theory. This framework assumes thin plate limit ($t \ll R$, with membrane thickness $t \approx 0.7$~nm and typical curvature radii $R \gtrsim 30$~nm), small slopes ($|\nabla z| \ll 1$), linear isotropic elasticity, and bending-dominated deformation.
The starting point is the Hessian matrix of the topography, which encodes the local curvature tensor:
\begin{equation}
\label{eq:SI_hessian}
H(z)=
\begin{pmatrix}
\frac{\partial^{2} z}{\partial x^{2}} & \frac{\partial^{2} z}{\partial x\,\partial y} \\[4pt]
\frac{\partial^{2} z}{\partial y\,\partial x} & \frac{\partial^{2} z}{\partial y^{2}}
\end{pmatrix}
=
\begin{pmatrix}
h_{xx} & h_{xy} \\
h_{xy} & h_{yy}
\end{pmatrix}
\end{equation}
From $H(z)$ we construct the mean curvature $\omega=\tfrac{1}{2}(h_{xx}+h_{yy})$ and the Gaussian curvature $G=h_{xx}h_{yy}-h_{xy}h_{yx}$. The principal curvatures, which represent the maximum and minimum curvatures along orthogonal directions, are the eigenvalues of $H$:
\begin{equation}
\kappa_{1,2}=\omega \pm \sqrt{\omega^{2}-G}
\end{equation}
Within the Kirchhoff--Love framework, a membrane of thickness $t$ bent with principal curvatures $\kappa_1$ and $\kappa_2$ experiences a surface bending-strain magnitude:
\begin{equation}
\label{eq:SI_full_strain}
\varepsilon(x,y)=\frac{t}{2}\,\sqrt{\kappa_{1}^{2}+\kappa_{2}^{2}}
\end{equation}
Analysis of our AFM maps reveals that $h_{xx} \sim h_{yy}$ and $h_{xy} \sim h_{yx}$, with $h_{xx}, h_{yy} \gg h_{xy},h_{yx} $, leading to $|G| \ll \omega^{2}$ over most suspended regions, indicating cylindrical bending where one principal curvature dominates while the other is negligible: $|\kappa_1| \gg |\kappa_2| \approx 0$. In this regime, $\kappa_1 \simeq 2\omega$ and $\kappa_2 \simeq 0$, reducing Eq.~\eqref{eq:SI_full_strain} to the simplified form presented in the main text:
\begin{equation}
\label{eq:SI_simplified}
\varepsilon(x,y) \approx t|\omega| = \frac{t}{2}\left|\partial_x^2 z + \partial_y^2 z\right|
\end{equation}

\begin{figure}[H]
    \centering
    \includegraphics[width=0.5\linewidth]{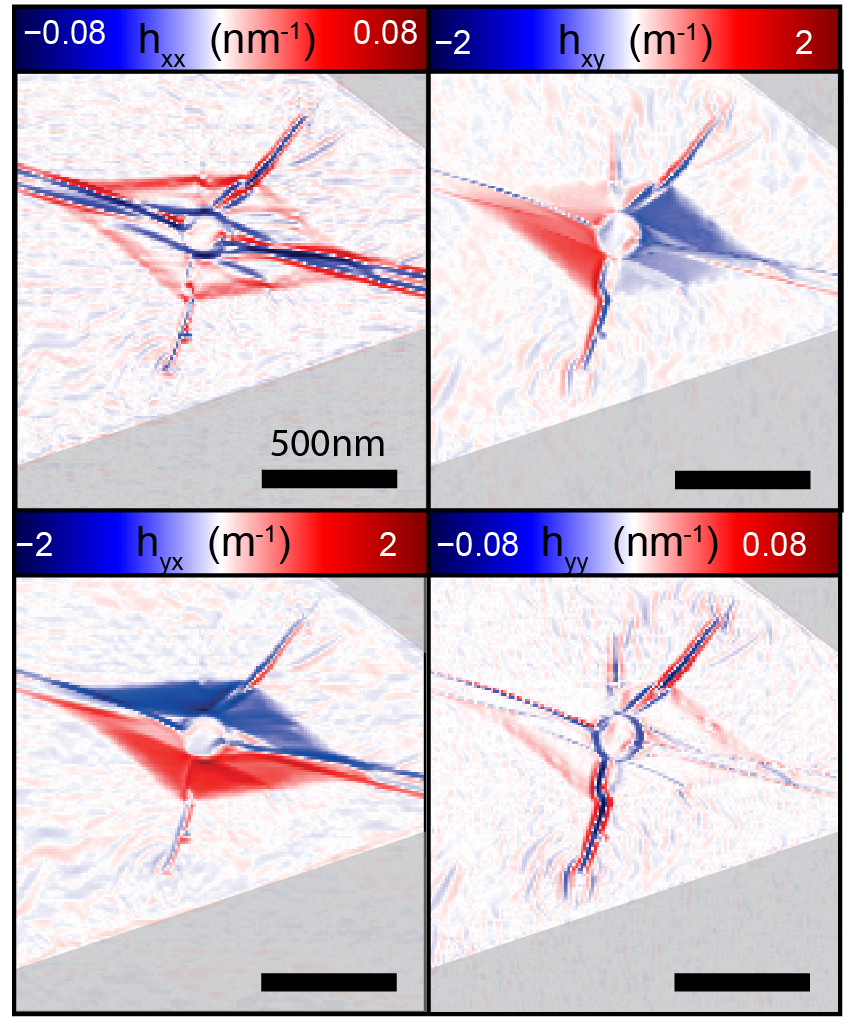}
    \caption{\textbf{Components of the topography Hessian matrix.} 
Spatial maps of $h_{xx}$, $h_{yy}$, and $h_{xy}$ extracted from AFM measurements. The diagonal terms dominate and control the mean curvature, while the off-diagonal terms $h_{xy}$ and $h_{yx}$ remain small except at tent facets where two-directional bending occurs. Color scales are independently adjusted for each panel to enhance contrast. Scale bar: 500~nm.}
\label{fig:SI_hessian_components}
\end{figure}

Figure~\ref{fig:SI_hessian_components} displays the individual components of the Hessian matrix (Eq. ~\ref{eq:SI_hessian} across a representative region. The off-diagonal term $h_{xy}$ is negligible compared to the diagonal terms $h_{xx}$ and $h_{yy}$ that govern the mean curvature. Nonetheless, the off-diagonal component shows spatial consistency with the expected Gaussian curvature distribution: enhanced $|h_{xy}|$ values appear along the tent facets where the membrane bends in both $x$ and $y$ directions simultaneously, corresponding to saddle-like geometries around the pillar apex and at ripple intersections. This spatial correlation validates our curvature extraction procedure while confirming that such two-directional bending remains a localized feature that does not affect the overall strain distribution governed by the dominant principal curvature.

\FloatBarrier

\section{Role of background noise and emitter intensity fluctuations in super-localization precision}
\label{si:superloc-precision}

Super-localization methods rely on extracting the coordinates of the centroid of a Gaussian emission spot. The precision of position determination is governed by two main factors: (i) the signal-to-noise ratio of the extracted emission profile, which depends on how well the emission peak is spectrally isolated and how efficiently background noise can be rejected, and (ii) physical effects intrinsic to the emitter itself, such as spectral wandering, intensity jittering, or blinking, which can affect the collected signal.
We discuss here how these effects impact the precision of the extracted coordinates through two complementary numerical simulations.

\begin{figure*}[htbp]
\centering
\includegraphics[width=\textwidth]{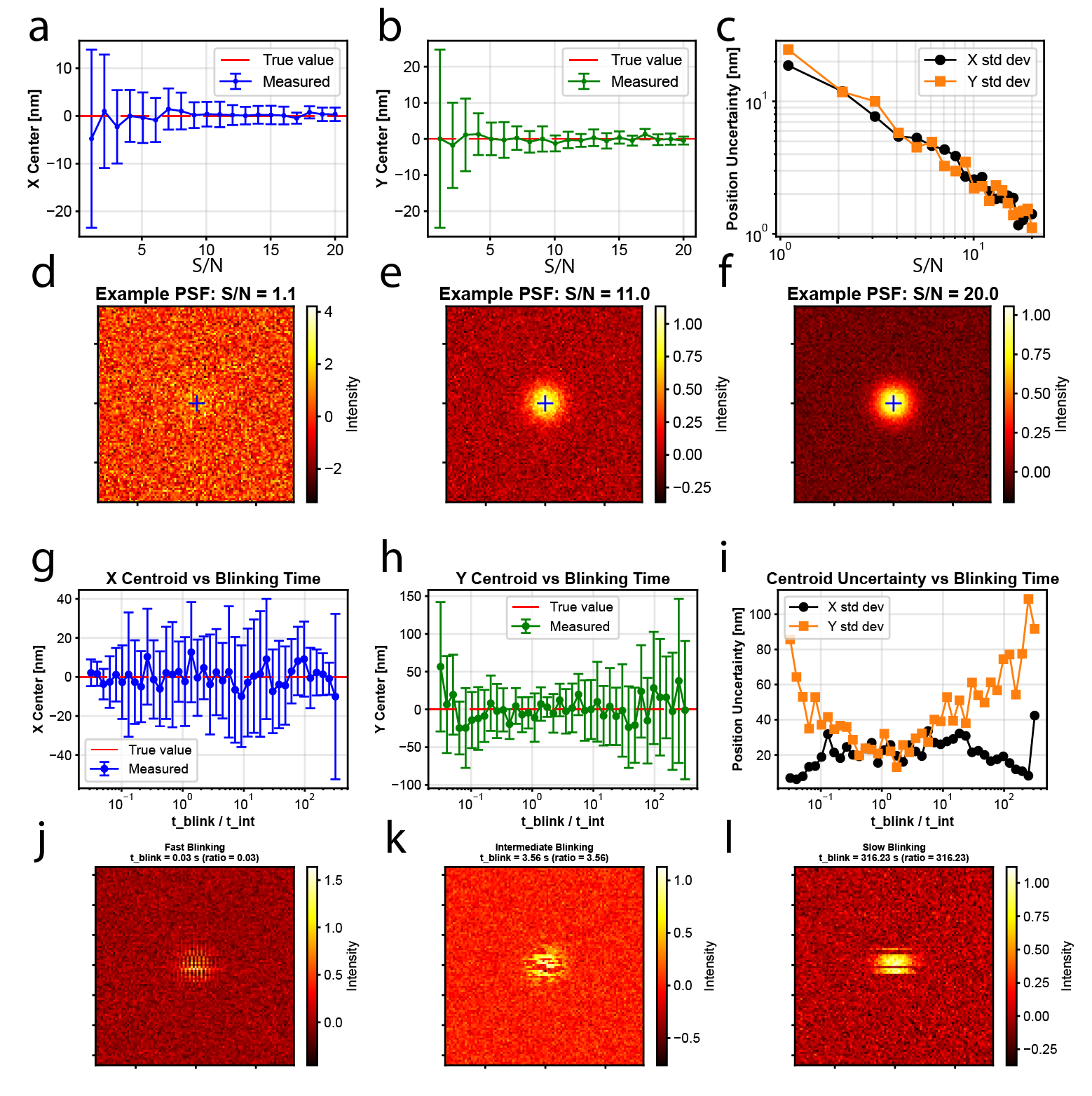}
\caption{\textbf{Analysis of signal-to-noise ratio and blinking effects on centroid localization precision.} 
\textbf{(a-c)}: Centroid position along X and Y axes and position uncertainty versus S/N ratio (1.1 to 20) for a simulated diffraction-limited PSF (FWHM = 500 nm). Error bars represent the standard deviation over 20 trials. Red dashed lines indicate the true center (0 nm). Log-log plot (c) shows an inverse relationship between S/N and localization uncertainty. 
\textbf{(d-f)}: Example PSF realizations at S/N = 1.1, 10.5, and 20.0, respectively. 
\textbf{(g-i)}: Centroid position and uncertainty versus blinking time ratio ($t_{\text{blink}}/t_{\text{int}}$) during raster scanning ($t_{\text{int}}$ = 1 s/pixel, S/N = 10, 15 trials per ratio). X-coordinate (g) shows weak dependence on blinking, while Y-coordinate (h) exhibits degradation at extreme ratios due to temporal correlation along the slow-scan axis. Panel (i) shows anisotropic uncertainty behavior. 
\textbf{(j-l)}: PSF examples for fast, intermediate, and slow blinking regimes, showing horizontal striations, optimal Gaussian appearance, and vertical banding, respectively. Blue crosses mark true center positions. Colormaps: hot (black to white/yellow = increasing intensity).}
\label{fig:S1}
\end{figure*}

The first simulation examines the fundamental role of signal-to-noise ratio on centroid determination. We simulate a diffraction-limited point spread function (PSF) with FWHM = 500 nm and perform Gaussian fitting for multiple S/N ratios ranging from 1.1 to 20, with 20 trials per S/N value.
Figures ~\ref{fig:S1}a-c show the extracted centroid position and uncertainties along both X and Y directions. As expected, higher S/N ratios yield smaller positional uncertainties in both directions, while the mean centroid position remains unbiased regardless of noise level. Panels d-f illustrate representative PSF realizations at low (S/N = 1.1), intermediate (S/N = 10.5), and high (S/N = 20.0) signal-to-noise ratios, showing the progressive improvement in emission spot quality.

An individual quantum emitter may exhibit stochastic blinking, randomly switching between bright (ON) and dark (OFF) states. This noise degrades localization accuracy beyond that caused by simple Gaussian noise.

As an example, let's consider a blinking dynamics governed by a two-state Poisson process. When the emitter is in a given state (ON or OFF), it remains in that state for a random duration $t_{\text{wait}}$ before switching to the other state. This waiting time follows an exponential distribution:
\[
p(t) = \frac{1}{t_{\text{blink}}} e^{-t/t_{\text{blink}}}
\]
where $t_{\text{blink}}$ is the characteristic blinking timescale. Physically, $t_{\text{blink}}$ represents the average time the emitter spends in each state before switching. For example, if $t_{\text{blink}} = 1$ s, the emitter will, on average, stay ON for 1 second, then switch OFF for another $\sim$1 second on average, and so on. However, each individual waiting time $t_{\text{wait}}$ is random, i.e., sometimes shorter, sometimes longer than $t_{\text{blink}}$.

In our experiment, a typical confocal (hyperspectral) image is acquired via line-by-line raster scanning from bottom to top. Each pixel is measured for a fixed integration time $t_{\text{int}}$ (typically 10~s). During this integration period, the detector accumulates all photons arriving from the emitter. Crucially, the emitter's blinking state evolves continuously throughout the scan—it does not reset between pixels. The emitter may blink multiple times during a single pixel integration ($t_{\text{wait}} < t_{\text{int}}$), or it may remain in the same state across many consecutive pixels ($t_{\text{wait}} > t_{\text{int}}$).

For a pixel at position $(i,j)$ in an image of width $N$ pixels (with $i$ indexing the line number and $j$ the position within the line, both starting from 0), the measurement begins at time $t = (Ni + j) t_{\text{int}}$. The recorded signal is:
\[
S(i,j) = \text{PSF}_{\text{0}}(i,j) \, f_{\text{ON}}(i,j) + \mathcal{N}(0, \sigma_{\text{noise}})
\]
where $\text{PSF}_{\text{0}}(i,j)$ is the ideal Gaussian point spread function, $f_{\text{ON}}(i,j) \in [0, 1]$ is the fraction of time the emitter spent in the ON state during that pixel's integration period, and $\mathcal{N}(0, \sigma_{\text{noise}})$ is additive Gaussian noise.

The key dimensionless parameter determining the blinking regime and its impact on the PSF is the ratio $t_{\text{blink}}/t_{\text{int}}$. When $t_{\text{blink}}/t_{\text{int}} \ll 1$ (fast blinking), the emitter switches states many times during each pixel integration. Each pixel samples multiple ON/OFF cycles, resulting in different random $f_{\text{ON}}$ values across pixels and creating spatial heterogeneity. When $t_{\text{blink}}/t_{\text{int}} \sim 1$ (intermediate regime), on average, about one state transition occurs per pixel integration. This provides good averaging, and the PSF resembles a noisy Gaussian spot. When $t_{\text{blink}}/t_{\text{int}} \gg 1$ (slow blinking), the emitter rarely switches states during the scan. It may remain ON or OFF for many consecutive pixels, creating large-scale spatial patterns in the image. This dimensionless ratio $t_{\text{blink}}/t_{\text{int}}$ is therefore the natural parameter for quantifying blinking effects in scanning measurements, as it captures the relative timescales of the emitter's intrinsic dynamics and the measurement time.

Our second simulation thus explores this blinking effect by testing blinking timescales ranging from 0.1 s to 100 s (corresponding to $t_{\text{blink}}/t_{\text{int}}$ from 0.1 to 100) for a fixed integration time of $t_{\text{int}} = 1$ s per pixel. We perform 15 Monte Carlo trials per blinking time and fit each realization with a 2D Gaussian.
Figure ~\ref{fig:S1}g-i reveals the anisotropic impact of blinking on localization precision. Since our raster scan proceeds along the X (fast) axis, followed by the Y (slow) axis, the X-coordinate is relatively insensitive to the blinking ratio. In contrast, the Y-coordinate shows strong degradation in both extreme regimes. This anisotropy arises from the temporal correlation of the blinking process: along the fast-scanning X direction, each line is acquired quickly, averaging over multiple blink events, whereas along the slow Y direction, the emitter's state can remain correlated across multiple lines. Representative PSF realizations (panels j-l) illustrate three distinct regimes:
\begin{itemize}
    \item \textbf{Fast blinking} ($t_{\text{blink}}/t_{\text{int}} \ll 1$, panel j): The emission spot exhibits vertical striations, as the emitter rapidly switches states within each scan line, creating spatial heterogeneity along X.
    \item \textbf{Intermediate blinking} ($t_{\text{blink}}/t_{\text{int}} \sim 1$, panel k): The PSF resembles a simple noisy Gaussian spot, yielding the best fit quality and smallest localization errors.
    \item \textbf{Slow blinking} ($t_{\text{blink}}/t_{\text{int}} \gg 1$, panel l): Horizontal banding appears, as the emitter remains in one state for extended periods, affecting multiple consecutive scan lines. Figure~\ref{fig:fig4}c illustrates this behavior: the intensity profile for $I_3$ exhibits horizontal bands with abrupt intensity jumps between scan lines, consistent with slow blinking where the emitter switches states infrequently but remains partially active throughout the measurement.
\end{itemize}
Note that at even slower timescales ($t_{\text{blink}} \gtrsim t_{\text{int}} \times N_{\text{pixels}}$), the emitter remains in a single state throughout the scan, effectively eliminating observable blinking—the map is then either uniformly bright or dark.
Both extreme regimes degrade the fit quality by breaking the assumed circular symmetry of the Gaussian PSF model, introducing systematic errors and increased uncertainty. 
In our experiments, typical integration times are approximately 10 s per pixel. The emission spots we observe are consistent with weak blinking effects (large $t_{\text{blink}}/t_{\text{int}}$ ratios) and high signal-to-noise ratios (S/N $>$ 20).

\section{Registration of AFM and PL Coordinate Systems}
\label{si:coordinate-system}

To correlate the positions of localized states (LS) extracted from hyperspectral $\mu$-PL superlocalization with AFM-derived strain and phase maps, it was necessary to register both datasets within a common coordinate system. This procedure involved defining consistent origins, extracting angular orientations, and applying rigid-body transformations (translation and rotation), while also evaluating possible distortions in the PL maps.

In the PL reference frame, the origin was defined as the center of a selected nanopillar. This point was obtained from the hyperspectral PL map by fitting a two-dimensional Gaussian profile to the intensity distribution. The angular orientation of the PL coordinate system was then determined from reflection maps: the centers of several aligned pillars were manually identified, and a linear regression through these points provided the orientation angle relative to the horizontal axis.

In the AFM maps, the pillar center was determined by applying a Canny edge-detection filter to the topography, followed by a circular fit to the extracted contour. This yielded a precise geometrical center, which was set as the AFM origin. To define the angular orientation, multiple pillar centers were similarly obtained via circle fitting, and the resulting alignment line provided the AFM flake orientation.

\begin{figure}[h!]
    \centering
    \includegraphics[width=0.95\linewidth]{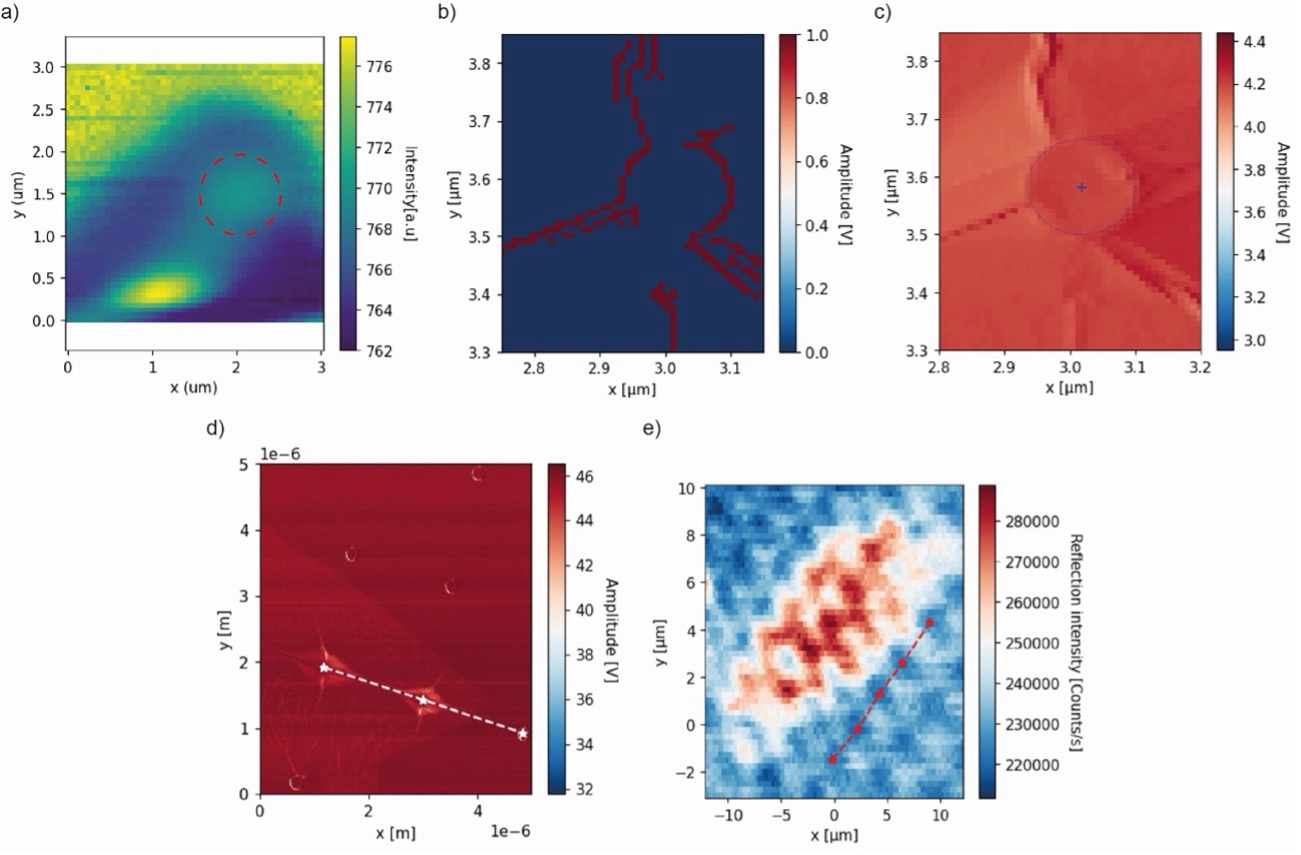}
    \caption{\textbf{Analysis of pillar positioning and flake orientation.} 
    \textbf{a}: Weighted PL intensity map with the top pillar marked (red dashed circle). 
    \textbf{b}: Pillar edges extracted from an amplitude map using a Canny filter (sigma = 0.2, thresholds = 0.1/0.2).  
    \textbf{c}: Circular fit of the pillar edge, giving a diameter of 165~nm and center at (3.02, 3.58) µm in AFM space.
    \textbf{d}: Amplitude map of three pillars used to determine flake orientation (\SI{344.92}{\degree}).  
    \textbf{e}: Reflection map showing flake orientation (\SI{32.49}{\degree}).}
    \label{fig:figS1}
\end{figure}

The mapping between PL and AFM frames was achieved by applying a translation operator (shifting the PL origin to the AFM origin) followed by a rotation operator (defined by the difference in orientation angles). Applying these operators to the emitter coordinates in PL space yielded their positions in AFM coordinates. This alignment procedure is illustrated in Fig.~\ref{fig:figS1}.

\FloatBarrier

Potential distortions of the PL coordinate system were also considered. A stretching operator was tested in two scenarios: (i) anisotropic scaling along one axis (ratio $\sim$1.28), and (ii) deviation from orthogonality (basis angle $\sim$100$^\circ$). In both cases, the induced displacements were on the order of $\sim$2~$\mu$m, while the radial distances of emitters from the pillar center remained within the repositioning uncertainty of the translation stage. Only one emitter, located far from the center, showed a larger deviation, consistent with enhanced sensitivity to small rotational errors. These tests indicate that distortions in the PL maps do not significantly affect the registration accuracy.

To quantify the registration between AFM and PL coordinate systems, we analyzed a set of emitters measured repeatedly under different transformation assumptions. Table~S1 summarizes the comparison between PL-extracted coordinates and their AFM-registered counterparts. For each localized state (LS), we report the radial distance to the pillar center as obtained in PL coordinates, the corresponding AFM-registered value, and the residual deviation $\Delta r$.

\begin{table}[h!]
\centering
\caption{Comparison of several LS positions in PL and AFM reference frames. $r_{\mathrm{PL}}$ is the radial distance from the pillar center in the PL map, $r_{\mathrm{AFM}}$ is the value after registration into AFM coordinates, and $\Delta r = r_{\mathrm{PL}} - r_{\mathrm{AFM}}$.}
\begin{tabular}{lccc}
\toprule
Emitter ID & $r_{\mathrm{PL}}$ ($\mu$m) & $r_{\mathrm{AFM}}$ ($\mu$m) & $\Delta r$ ($\mu$m) \\
\midrule
LS774 & 0.72 & 0.69 & 0.03 \\
LS775 & 0.85 & 0.81 & 0.04 \\
LS776 & 1.02 & 0.98 & 0.04 \\
LS777 & 1.10 & 1.13 & -0.03 \\
LS778 & 1.25 & 1.21 & 0.04 \\
LS779 & 1.30 & 1.28 & 0.02 \\
LS780 & 1.40 & 1.42 & -0.02 \\
\bottomrule
\end{tabular}
\label{tab:registration}
\end{table}

The data show that the typical deviation between PL and AFM coordinates is below 50~nm, well within the uncertainty associated with the stage repositioning and the Gaussian fitting error in the superlocalization procedure. Only one emitter located farther from the pillar center displayed a deviation close to 100~nm, which we attribute to enhanced sensitivity to small angular misalignments at larger radial distances.  

This quantitative analysis confirms that the registration procedure introduces negligible systematic error compared to the intrinsic resolution of our PL superlocalization maps.

\section{Statistical Analysis: Random vs. Strain-Threshold Distribution}
\label{si:stat_analysis}
To assess whether quantum emitter localization is influenced by local strain, we performed Monte Carlo simulations comparing the observed spatial distribution against a null hypothesis of random placement. We generated 1,000 independent realizations of 14 randomly positioned emitters within the measured region (2.72--3.27~$\mu$m $\times$ 1.15--1.69~$\mu$m), accounting for the 50~nm position uncertainty.
For each simulation, we calculated the number of emitters whose uncertainty circles intersect high-strain regions defined by various percentile thresholds. The $p$-value is the fraction of Monte Carlo runs that yield overlap at least as large as that observed in the experiment, quantifying the probability of obtaining the observed result under a random distribution.

\begin{table}[h]
\centering
\caption{Strain threshold analysis showing observed vs. expected emitter overlap with high-strain regions. Enrichment factor $E = N_{\text{obs}}/N_{\text{exp}}$. Statistical significance achieved at the 90th percentile ($p < 0.05$).}
\label{tab:strain_threshold}
\begin{tabular}{cccccc}
\hline
\textbf{Percentile} & \textbf{Threshold} & \textbf{$N_{\text{obs}}$} & \textbf{$N_{\text{exp}}$} & \textbf{Enrichment} & \textbf{$p$-value} \\
\textbf{(\%)} & \textbf{($\varepsilon$, \%)} & & & \textbf{($E$)} & \\
\hline
10 & 0.066 & 14 & 14.0 $\pm$ 0.0 & 1.00 & 1.000 \\
20 & 0.111 & 14 & 14.0 $\pm$ 0.1 & 1.00 & 0.998 \\
30 & 0.157 & 14 & 13.8 $\pm$ 0.4 & 1.02 & 0.782 \\
40 & 0.235 & 14 & 12.7 $\pm$ 0.9 & 1.10 & 0.253 \\
50 & 0.355 & 11 & 12.1 $\pm$ 1.2 & 0.91 & 0.892 \\
60 & 0.494 & 11 & 11.2 $\pm$ 1.4 & 0.99 & 0.709 \\
70 & 0.674 & 10 & 9.9 $\pm$ 1.6 & 1.01 & 0.608 \\
80 & 0.949 & 10 & 7.9 $\pm$ 1.9 & 1.26 & 0.201 \\
\textbf{90} & \textbf{1.472} & \textbf{9} & \textbf{5.1 $\pm$ 1.8} & \textbf{1.76} & \textbf{0.035} \\
\hline
\end{tabular}
\end{table}

The results show that enrichment values $E$ increase with increasing strain threshold, reaching statistical significance at the 90th percentile ($\varepsilon > 1.47\%$), with $E = 1.76$ and $p = 0.035 < 0.05$. This indicates emitters preferentially occupy high-strain regions with $\sim$1.8$\times$ enhanced probability compared to random distribution. However, the moderate enrichment factor and presence of low-strain emitters (5 out of 14 below 0.5\% strain) demonstrate that strain is a significant but non-exclusive localization factor, consistent with a multi-pathway formation mechanism involving both strain and dielectric effects.

\begin{acknowledgments}
The authors thank François Dubin, Joël Moser and Fabien Vialla for insightful discussions.
A.R.-P. acknowledges funding from ANR JCJC NEAR-2D (ANR-23-CE47-0015), Idex Welcome Package (UniCA), Fédération Doeblin FR2800, and CNRS Physique. This work was supported by the French government through the UCA-JEDI Investments in the Future project managed by the National Research Agency (ANR-15-IDEX-01).
\end{acknowledgments}


\begin{thebibliography}{99}
\bibitem{agarwal_situ_2025} Agarwal, Hitesh, Reserbat-Plantey, Antoine, Ruiz, David Barcons, Soundarapandian, Karuppasammy, Li, Geng, Mkhitaryan, Vahagn, Osmond, Johann, Lozano, Helena, Watanabe, Kenji, Taniguchi, Takashi, Stepanov, Petr, Koppens, Frank H. L., Kumar, Roshan Krishna, ``In situ engineering hexagonal boron nitride in van der {Waals} heterostructures with selective {SF6} etching'', arXiv, (2025), doi:10.48550/arXiv.2503.14238.

\bibitem{palacios-berraquero_large-scale_2017} Palacios-Berraquero, Carmen, Kara, Dhiren M., Montblanch, Alejandro R.-P., Barbone, Matteo, Latawiec, Pawel, Yoon, Duhee, Ott, Anna K., Loncar, Marko, Ferrari, Andrea C., Atat{\"u}re, Mete, ``Large-scale quantum-emitter arrays in atomically thin semiconductors'', Nature Communications, \textbf{8}(1), 15093, (2017), doi:10.1038/ncomms15093.

\bibitem{aharonovich_solid-state_2016} Aharonovich, Igor, Englund, Dirk, Toth, Milos, ``Solid-state single-photon emitters'', Nature Photonics, \textbf{10}(10), 631--641, (2016), doi:10.1038/nphoton.2016.186.

\bibitem{azzam_prospects_2021} Azzam, Shaimaa I., Parto, Kamyar, Moody, Galan, ``Prospects and challenges of quantum emitters in {2D} materials'', Applied Physics Letters, \textbf{118}(24), 240502, (2021), doi:10.1063/5.0054116.

\bibitem{zhao_single_2018} Zhao, Shen, Lavie, Julien, Rondin, Lo{\"i}c, Orcin-Chaix, Lucile, Diederichs, Carole, Roussignol, Philippe, Chassagneux, Yannick, Voisin, Christophe, M{\"u}llen, Klaus, Narita, Akimitsu, Campidelli, St{\'e}phane, Lauret, Jean-S{\'e}bastien, ``Single photon emission from graphene quantum dots at room temperature'', Nature Communications, \textbf{9}(1), 3470, (2018), doi:10.1038/s41467-018-05888-w.

\bibitem{toninelli_single_2021} Toninelli, C., Gerhardt, I., Clark, A. S., Reserbat-Plantey, A., G{\"o}tzinger, S., Ristanovi{\'c}, Z., Colautti, M., Lombardi, P., Major, K. D., Deperasi{\'n}ska, I., Pernice, W. H., Koppens, F. H. L., Kozankiewicz, B., Gourdon, A., Sandoghdar, V., Orrit, M., ``Single organic molecules for photonic quantum technologies'', Nature Materials, \textbf{20}(12), 1615--1628, (2021), doi:10.1038/s41563-021-00987-4.

\bibitem{he_carbon_2018} He, X., Htoon, H., Doorn, S. K., Pernice, W. H. P., Pyatkov, F., Krupke, R., Jeantet, A., Chassagneux, Y., Voisin, C., ``Carbon nanotubes as emerging quantum-light sources'', Nature Materials, \textbf{17}(8), 663--670, (2018), doi:10.1038/s41563-018-0109-2.

\bibitem{tonndorf_single-photon_2015} Tonndorf, Philipp, Schmidt, Robert, Schneider, Robert, Kern, Johannes, Buscema, Michele, Steele, Gary A., Castellanos-Gomez, Andres, Zant, Herre S. J. van der, Vasconcellos, Steffen Michaelis de, Bratschitsch, Rudolf, ``Single-photon emission from localized excitons in an atomically thin semiconductor'', Optica, \textbf{2}(4), 347--352, (2015), doi:10.1364/OPTICA.2.000347.

\bibitem{srivastava_optically_2015} Srivastava, Ajit, Sidler, Meinrad, Allain, Adrien V., Lembke, Dominik S., Kis, Andras, Imamo{\u g}lu, A., ``Optically active quantum dots in monolayer {WSe2}'', Nature Nanotechnology, \textbf{10}(6), 491--496, (2015), doi:10.1038/nnano.2015.60.

\bibitem{koperski_single_2015} Koperski, M., Nogajewski, K., Arora, A., Cherkez, V., Mallet, P., Veuillen, J.-Y., Marcus, J., Kossacki, P., Potemski, M., ``Single photon emitters in exfoliated {WSe2} structures'', Nature Nanotechnology, \textbf{10}(6), 503--506, (2015), doi:10.1038/nnano.2015.67.

\bibitem{he_single_2015} He, Yu-Ming, Clark, Genevieve, Schaibley, John R., He, Yu, Chen, Ming-Cheng, Wei, Yu-Jia, Ding, Xing, Zhang, Qiang, Yao, Wang, Xu, Xiaodong, Lu, Chao-Yang, Pan, Jian-Wei, ``Single quantum emitters in monolayer semiconductors'', Nature Nanotechnology, \textbf{10}(6), 497--502, (2015), doi:10.1038/nnano.2015.75.

\bibitem{rosenberger_quantum_2019} Rosenberger, Matthew R., Dass, Chandriker Kavir, Chuang, Hsun-Jen, Sivaram, Saujan V., McCreary, Kathleen M., Hendrickson, Joshua R., Jonker, Berend T., ``Quantum {Calligraphy}: {Writing} {Single}-{Photon} {Emitters} in a {Two}-{Dimensional} {Materials} {Platform}'', ACS Nano, \textbf{13}(1), 904--912, (2019), doi:10.1021/acsnano.8b08730.

\bibitem{chakraborty_voltage-controlled_2015} Chakraborty, Chitraleema, Kinnischtzke, Laura, Goodfellow, Kenneth M., Beams, Ryan, Vamivakas, A. Nick, ``Voltage-controlled quantum light from an atomically thin semiconductor'', Nature Nanotechnology, \textbf{10}(6), 507--511, (2015), doi:10.1038/nnano.2015.79.

\bibitem{heithoff_valley-hybridized_2024} Heithoff, Maximilian, Moreno, {\'A}lvaro, Torre, Iacopo, Feuer, Matthew S. G., Purser, Carola M., Andolina, Gian Marcello, Calaj{\`o}, Giuseppe, Watanabe, Kenji, Taniguchi, Takashi, Kara, Dhiren M., Hays, Patrick, Tongay, Seth Ariel, Fal'ko, Vladimir I., Chang, Darrick, Atat{\"u}re, Mete, Reserbat-Plantey, Antoine, Koppens, Frank H.L., ``Valley-{Hybridized} {Gate}-{Tunable} {1D} {Exciton} {Confinement} in {MoSe2}'', ACS Nano, \textbf{18}(44), 30283--30292, (2024), doi:10.1021/acsnano.4c04786.

\bibitem{thureja_electrically_2022} Thureja, Deepankur, Imamoglu, Atac, Smole{\'n}ski, Tomasz, Amelio, Ivan, Popert, Alexander, Chervy, Thibault, Lu, Xiaobo, Liu, Song, Barmak, Katayun, Watanabe, Kenji, Taniguchi, Takashi, Norris, David J., Kroner, Martin, Murthy, Puneet A., ``Electrically tunable quantum confinement of neutral excitons'', Nature, \textbf{606}(7913), 298--304, (2022), doi:10.1038/s41586-022-04634-z.

\bibitem{thureja_electrically_2024} Thureja, Deepankur, Yaz{\i}c{\i}, F. Emre, Smole{\'n}ski, Tomasz, Kroner, Martin, Norris, David J., {\.I}mamo{\v g}lu, Atac, ``Electrically defined quantum dots for bosonic excitons'', Physical Review B, \textbf{110}(24), 245425, (2024), doi:10.1103/PhysRevB.110.245425.

\bibitem{baek_highly_2020} Baek, H., Brotons-Gisbert, M., Koong, Z. X., Campbell, A., Rambach, M., Watanabe, K., Taniguchi, T., Gerardot, B. D., ``Highly energy-tunable quantum light from moir{\'e}-trapped excitons'', Science Advances, \textbf{6}(37), eaba8526, (2020), doi:10.1126/sciadv.aba8526.

\bibitem{seyler_signatures_2019} Seyler, Kyle L., Rivera, Pasqual, Yu, Hongyi, Wilson, Nathan P., Ray, Essance L., Mandrus, David G., Yan, Jiaqiang, Yao, Wang, Xu, Xiaodong, ``Signatures of moir{\'e}-trapped valley excitons in {MoSe2}/{WSe2} heterobilayers'', Nature, \textbf{567}(7746), 66--70, (2019), doi:10.1038/s41586-019-0957-1.

\bibitem{branny_deterministic_2017} Branny, Artur, Kumar, Santosh, Proux, Rapha{\"e}l, Gerardot, Brian D., ``Deterministic strain-induced arrays of quantum emitters in a two-dimensional semiconductor'', Nature Communications, \textbf{8}(1), 15053, (2017), doi:10.1038/ncomms15053.

\bibitem{kim_position_2019} Kim, Hyoju, Moon, Jong Sung, Noh, Gichang, Lee, Jieun, Kim, Je-Hyung, ``Position and {Frequency} {Control} of {Strain}-{Induced} {Quantum} {Emitters} in {WSe}$_{\textrm{2}}$ {Monolayers}'', Nano Letters, \textbf{19}(10), 7534--7539, (2019), doi:10.1021/acs.nanolett.9b03421.

\bibitem{luo_deterministic_2018} Luo, Yue, Shepard, Gabriella D., Ardelean, Jenny V., Rhodes, Daniel A., Kim, Bumho, Barmak, Katayun, Hone, James C., Strauf, Stefan, ``Deterministic coupling of site-controlled quantum emitters in monolayer {WSe}$_{\textrm{2}}$ to plasmonic nanocavities'', Nature Nanotechnology, \textbf{13}(12), 1137--1142, (2018), doi:10.1038/s41565-018-0275-z.

\bibitem{linhart_localized_2019} Linhart, Lukas, Paur, Matthias, Smejkal, Valerie, Burgd{\"o}rfer, Joachim, Mueller, Thomas, Libisch, Florian, ``Localized {Intervalley} {Defect} {Excitons} as {Single}-{Photon} {Emitters} in \$\{{\textbackslash}mathrm\{{WSe}\}\}\_\{2\}\$'', Physical Review Letters, \textbf{123}(14), 146401, (2019), doi:10.1103/PhysRevLett.123.146401.

\bibitem{kern_nanoscale_2016} Kern, Johannes, Niehues, Iris, Tonndorf, Philipp, Schmidt, Robert, Wigger, Daniel, Schneider, Robert, Stiehm, Torsten, Michaelis de Vasconcellos, Steffen, Reiter, Doris E., Kuhn, Tilmann, Bratschitsch, Rudolf, ``Nanoscale {Positioning} of {Single}-{Photon} {Emitters} in {Atomically} {Thin} {WSe2}'', Advanced Materials, \textbf{28}(33), 7101--7105, (2016), doi:10.1002/adma.201600560.

\bibitem{fournier_position-controlled_2021} Fournier, Clarisse, Plaud, Alexandre, Roux, S{\'e}bastien, Pierret, Aur{\'e}lie, Rosticher, Michael, Watanabe, Kenji, Taniguchi, Takashi, Buil, St{\'e}phanie, Qu{\'e}lin, Xavier, Barjon, Julien, Hermier, Jean-Pierre, Delteil, Aymeric, ``Position-controlled quantum emitters with reproducible emission wavelength in hexagonal boron nitride'', Nature Communications, \textbf{12}(1), 3779, (2021), doi:10.1038/s41467-021-24019-6.

\bibitem{parto_defect_2021} Parto, Kamyar, Azzam, Shaimaa I., Banerjee, Kaustav, Moody, Galan, ``Defect and strain engineering of monolayer {WSe2} enables site-controlled single-photon emission up to 150 {K}'', Nature Communications, \textbf{12}(1), 3585, (2021), doi:10.1038/s41467-021-23709-5.

\bibitem{klein_site-selectively_2019} Klein, J., Lorke, M., Florian, M., Sigger, F., Sigl, L., Rey, S., Wierzbowski, J., Cerne, J., M{\"u}ller, K., Mitterreiter, E., Zimmermann, P., Taniguchi, T., Watanabe, K., Wurstbauer, U., Kaniber, M., Knap, M., Schmidt, R., Finley, J. J., Holleitner, A. W., ``Site-selectively generated photon emitters in monolayer {MoS2} via local helium ion irradiation'', Nature Communications, \textbf{10}(1), 2755, (2019), doi:10.1038/s41467-019-10632-z.

\bibitem{asenjo-garcia_exponential_2017} Asenjo-Garcia, A., Moreno-Cardoner, M., Albrecht, A., Kimble, H. J., Chang, D. E., ``Exponential {Improvement} in {Photon} {Storage} {Fidelities} {Using} {Subradiance} and "{Selective} {Radiance}" in {Atomic} {Arrays}'', Physical Review X, \textbf{7}(3), 031024, (2017), doi:10.1103/PhysRevX.7.031024.

\bibitem{wang_utilizing_2022} Wang, Wei, Jones, Leighton O., Chen, Jia-Shiang, Schatz, George C., Ma, Xuedan, ``Utilizing {Ultraviolet} {Photons} to {Generate} {Single}-{Photon} {Emitters} in {Semiconductor} {Monolayers}'', ACS Nano, \textbf{16}(12), 21240--21247, (2022), doi:10.1021/acsnano.2c09209.

\bibitem{raja_coulomb_2017} Raja, Archana, Chaves, Andrey, Yu, Jaeeun, Arefe, Ghidewon, Hill, Heather M., Rigosi, Albert F., Berkelbach, Timothy C., Nagler, Philipp, Sch{\"u}ller, Christian, Korn, Tobias, Nuckolls, Colin, Hone, James, Brus, Louis E., Heinz, Tony F., Reichman, David R., Chernikov, Alexey, ``Coulomb engineering of the bandgap and excitons in two-dimensional materials'', Nature Communications, \textbf{8}(1), 15251, (2017), doi:10.1038/ncomms15251.

\bibitem{ben_mhenni_breakdown_2025} Ben Mhenni, Amine, Van Tuan, Dinh, Geilen, Leonard, Petri{\'c}, Marko M., Erdi, Melike, Watanabe, Kenji, Taniguchi, Takashi, Tongay, Seth Ariel, M{\"u}ller, Kai, Wilson, Nathan P., Finley, Jonathan J., Dery, Hanan, Barbone, Matteo, ``Breakdown of the {Static} {Dielectric} {Screening} {Approximation} of {Coulomb} {Interactions} in {Atomically} {Thin} {Semiconductors}'', ACS Nano, \textbf{19}(4), 4269--4278, (2025), doi:10.1021/acsnano.4c11563.

\bibitem{itzhak_exciton_2025} Itzhak, Raziel, Suleymanov, Nathan, Minkovich, Boris, Kartvelishvili, Liana, Kostianovski, Vladislav, Korobko, Roman, Hayat, Alex, Goykhman, Ilya, ``Exciton {Manipulation} via {Dielectric} {Environment} {Engineering} in {2D} {Semiconductors}'', ACS Applied Optical Materials, \textbf{3}(6), 1330--1338, (2025), doi:10.1021/acsaom.5c00105.

\bibitem{noauthor_coupling_nodate} ``Coupling {Single} {Photons} from {Discrete} {Quantum} {Emitters} in {WSe2} to {Lithographically} {Defined} {Plasmonic} {Slot} {Waveguides} {\textbar} {Nano} {Letters}''.

\bibitem{blauth_coupling_2018} Blauth, M., J{\"u}rgensen, M., Vest, G., Hartwig, O., Prechtl, M., Cerne, J., Finley, J. J., Kaniber, M., ``Coupling {Single} {Photons} from {Discrete} {Quantum} {Emitters} in {WSe2} to {Lithographically} {Defined} {Plasmonic} {Slot} {Waveguides}'', Nano Letters, \textbf{18}(11), 6812--6819, (2018), doi:10.1021/acs.nanolett.8b02687.

\bibitem{xu_subdiffraction_2024} Xu, David D., Vong, Albert F., Utama, M. Iqbal Bakti, Lebedev, Dmitry, Ananth, Riddhi, Hersam, Mark C., Weiss, Emily A., Mirkin, Chad A., ``Sub-{Diffraction} {Correlation} of {Quantum} {Emitters} and {Local} {Strain} {Fields} in {Strain}-{Engineered} {WSe} 2 {Monolayers}'', Advanced Materials, \textbf{36}(25), (2024), doi:10.1002/adma.202314242.

\bibitem{abramov_photoluminescence_2023} Abramov, Artem N., Chestnov, Igor Y., Alimova, Ekaterina S., Ivanova, Tatiana, Mukhin, Ivan S., Krizhanovskii, Dmitry N., Shelykh, Ivan A., Iorsh, Ivan V., Kravtsov, Vasily, ``Photoluminescence imaging of single photon emitters within nanoscale strain profiles in monolayer {WSe2}'', Nature Communications, \textbf{14}(1), 5737, (2023), doi:10.1038/s41467-023-41292-9.

\bibitem{luo_imaging_2023} Luo, Weijun, Lawrie, Benjamin J., Puretzky, Alexander A., Tan, Qishuo, Gao, Hongze, Lingerfelt, David B., Eichman, Gage, Mcgee, Edward, Swan, Anna K., Liang, Liangbo, Ling, Xi, ``Imaging {Strain}-{Localized} {Single}-{Photon} {Emitters} in {Layered} {GaSe} below the {Diffraction} {Limit}'', ACS Nano, \textbf{17}(23), 23455--23465, (2023), doi:10.1021/acsnano.3c05250.

\bibitem{yu_site-controlled_2021} Yu, Leo, Deng, Minda, Zhang, Jingyuan Linda, Borghardt, Sven, Kardynal, Beata, Vu{\v c}kovi{\'c}, Jelena, Heinz, Tony F., ``Site-{Controlled} {Quantum} {Emitters} in {Monolayer} {MoSe2}'', Nano Letters, \textbf{21}(6), 2376--2381, (2021), doi:10.1021/acs.nanolett.0c04282.

\bibitem{kim_confinement_2024} Kim, Gwangwoo, Huet, Benjamin, Stevens, Christopher E., Jo, Kiyoung, Tsai, Jeng-Yuan, Bachu, Saiphaneendra, Leger, Meghan, Song, Seunguk, Rahaman, Mahfujur, Ma, Kyung Yeol, Glavin, Nicholas R., Shin, Hyeon Suk, Alem, Nasim, Yan, Qimin, Hendrickson, Joshua R., Redwing, Joan M., Jariwala, Deep, ``Confinement of excited states in two-dimensional, in-plane, quantum heterostructures'', Nature Communications, \textbf{15}(1), 6361, (2024), doi:10.1038/s41467-024-50653-x.

\bibitem{akinwande_graphene_2019} Akinwande, Deji, Huyghebaert, Cedric, Wang, Ching-Hua, Serna, Martha I., Goossens, Stijn, Li, Lain-Jong, Wong, H.-S. Philip, Koppens, Frank H. L., ``Graphene and two-dimensional materials for silicon technology'', Nature, \textbf{573}(7775), 507--518, (2019), doi:10.1038/s41586-019-1573-9.

\bibitem{hu_quantum_2024} Hu, Jenny, Lorchat, Etienne, Chen, Xueqi, Watanabe, Kenji, Taniguchi, Takashi, Heinz, Tony F., Murthy, Puneet A., Chervy, Thibault, ``Quantum control of exciton wave functions in {2D} semiconductors'', Science Advances, \textbf{10}(12), eadk6369, (2024), doi:10.1126/sciadv.adk6369.

\bibitem{barcons_ruiz_engineering_2022} Barcons Ruiz, David, Herzig Sheinfux, Hanan, Hoffmann, Rebecca, Torre, Iacopo, Agarwal, Hitesh, Kumar, Roshan Krishna, Vistoli, Lorenzo, Taniguchi, Takashi, Watanabe, Kenji, Bachtold, Adrian, Koppens, Frank H. L., ``Engineering high quality graphene superlattices via ion milled ultra-thin etching masks'', Nature Communications, \textbf{13}(1), 6926, (2022), doi:10.1038/s41467-022-34734-3.

\bibitem{soubelet_strong_2025} Soubelet, Pedro, Tong, Yao, Astaburuaga Hernandez, Asier, Ji, Peirui, Gallo, Katia, Stier, Andreas V., Finley, Jonathan J., ``Strong {Quantum} {Confinement} of {2D} {Excitons} in an {Engineered} {1D} {Potential} {Induced} by {Proximal} {Ferroelectric} {Domain} {Walls}'', Nano Letters, \textbf{25}(34), 12842--12850, (2025), doi:10.1021/acs.nanolett.5c02438.

\bibitem{hernandez_lopez_strain_2022} Hern{\'a}ndez L{\'o}pez, Pablo, Heeg, Sebastian, Schattauer, Christoph, Kovalchuk, Sviatoslav, Kumar, Abhijeet, Bock, Douglas J., Kirchhof, Jan N., H{\"o}fer, Bianca, Greben, Kyrylo, Yagodkin, Denis, Linhart, Lukas, Libisch, Florian, Bolotin, Kirill I., ``Strain control of hybridization between dark and localized excitons in a {2D} semiconductor'', Nature Communications, \textbf{13}(1), 7691, (2022), doi:10.1038/s41467-022-35352-9.

\bibitem{raynaud_superlocalization_2019} Raynaud, C., Claude, T., Borel, A., Amara, M. R., Graf, A., Zaumseil, J., Lauret, J.-S., Chassagneux, Y., Voisin, C., ``Superlocalization of {Excitons} in {Carbon} {Nanotubes} at {Cryogenic} {Temperature}'', Nano Letters, \textbf{19}(10), 7210--7216, (2019), doi:10.1021/acs.nanolett.9b02816.

\bibitem{Dirnberger2021} Dirnberger, Florian, Ziegler, Jonas D., Faria Junior, Paulo E., Bushati, Rezlind, Taniguchi, Takashi, Watanabe, Kenji, Fabian, Jaroslav, Bougeard, Dominique, Chernikov, Alexey, Menon, Vinod M., ``Quasi-{1D} exciton channels in strain-engineered {2D} materials'', Science Advances, \textbf{7}(44), (2021), doi:10.1126/sciadv.abj3066.

\bibitem{vasic_phase_2017} Vasi{\'c}, Borislav, Matkovi{\'c}, Aleksandar, Gaji{\'c}, Rado{\v s}, ``Phase imaging and nanoscale energy dissipation of supported graphene using amplitude modulation atomic force microscopy'', Nanotechnology, \textbf{28}(46), 465708, (2017), doi:10.1088/1361-6528/aa8e3b.

\bibitem{semond_gan_1999} Semond, F., Damilano, B., V{\'e}zian, S., Grandjean, N., Leroux, M., Massies, J., ``{GaN} grown on {Si}(111) substrate: {From} two-dimensional growth to quantum well assessment'', Applied Physics Letters, \textbf{75}(1), 82--84, (1999), doi:10.1063/1.124283.

\bibitem{noauthor_properties_nodate} ``Properties of {Advanced} {Semiconductor} {Materials}: {GaN}, {AIN}, {InN}, {BN}, {SiC}, {SiGe} {\textbar} {Wiley}'', Wiley.com.

\bibitem{levinstejn_properties_2001} Levin{\v s}tejn, Michail E., Rumyantsev, Sergey L., Shur, Michael, ``Properties of advanced semiconductor materials {GaN}, {AlN}, {InN}, {BN}, {SiC}, {SiGe}'', Wiley, (2001).

\bibitem{chernikov_exciton_2014} Chernikov, Alexey, Berkelbach, Timothy C., Hill, Heather M., Rigosi, Albert, Li, Yilei, Aslan, Burak, Reichman, David R., Hybertsen, Mark S., Heinz, Tony F., ``Exciton {Binding} {Energy} and {Nonhydrogenic} {Rydberg} {Series} in {Monolayer} \$\{{\textbackslash}mathrm\{{WS}\}\}\_\{2\}\$'', Physical Review Letters, \textbf{113}(7), 076802, (2014), doi:10.1103/PhysRevLett.113.076802.

\bibitem{xu_creation_2021} Xu, Yang, Horn, Connor, Zhu, Jiacheng, Tang, Yanhao, Ma, Liguo, Li, Lizhong, Liu, Song, Watanabe, Kenji, Taniguchi, Takashi, Hone, James C., Shan, Jie, Mak, Kin Fai, ``Creation of moir{\'e} bands in a monolayer semiconductor by spatially periodic dielectric screening'', Nature Materials, \textbf{20}(5), 645--649, (2021), doi:10.1038/s41563-020-00888-y.

\bibitem{van_tuan_effects_2024} Van Tuan, Dinh, Dery, Hanan, ``Effects of dynamical dielectric screening on the excitonic spectrum of monolayer semiconductors'', Physical Review B, \textbf{110}(12), 125301, (2024), doi:10.1103/PhysRevB.110.125301.

\bibitem{daveau_spectral_2020} Daveau, Rapha{\"e}l S., Vandekerckhove, Tom, Mukherjee, Arunabh, Wang, Zefang, Shan, Jie, Mak, Kin Fai, Vamivakas, A. Nick, Fuchs, Gregory D., ``Spectral and spatial isolation of single tungsten diselenide quantum emitters using hexagonal boron nitride wrinkles'', APL Photonics, \textbf{5}(9), 096105, (2020), doi:10.1063/5.0013825.

\bibitem{gelly_probing_2022} Gelly, Ryan J., Renaud, Dylan, Liao, Xing, Pingault, Benjamin, Bogdanovic, Stefan, Scuri, Giovanni, Watanabe, Kenji, Taniguchi, Takashi, Urbaszek, Bernhard, Park, Hongkun, Lon{\v c}ar, Marko, ``Probing dark exciton navigation through a local strain landscape in a {WSe2} monolayer'', Nature Communications, \textbf{13}(1), 232, (2022), doi:10.1038/s41467-021-27877-2.

\bibitem{tripathi_spontaneous_2018} Tripathi, Laxmi Narayan, Iff, Oliver, Betzold, Simon, Dusanowski, {\L}ukasz, Emmerling, Monika, Moon, Kihwan, Lee, Young Jin, Kwon, Soon-Hong, H{\"o}fling, Sven, Schneider, Christian, ``Spontaneous {Emission} {Enhancement} in {Strain}-{Induced} {WSe2} {Monolayer}-{Based} {Quantum} {Light} {Sources} on {Metallic} {Surfaces}'', ACS Photonics, \textbf{5}(5), 1919--1926, (2018), doi:10.1021/acsphotonics.7b01053.

\bibitem{campo_optical_2021} Campo, Jochen, Cambr{\'e}, Sofie, Botka, Bea, Obrzut, Jan, Wenseleers, Wim, Fagan, Jeffrey A., ``Optical {Property} {Tuning} of {Single}-{Wall} {Carbon} {Nanotubes} by {Endohedral} {Encapsulation} of a {Wide} {Variety} of {Dielectric} {Molecules}'', ACS Nano, \textbf{15}(2), 2301--2317, (2021), doi:10.1021/acsnano.0c08352.

\bibitem{walsh_scaling_2008} Walsh, Andrew G., Nickolas Vamivakas, A., Yin, Yan, Cronin, Stephen B., Selim {\"U}nl{\"u}, M., Goldberg, Bennett B., Swan, Anna K., ``Scaling of exciton binding energy with external dielectric function in carbon nanotubes'', Physica E: Low-dimensional Systems and Nanostructures, \textbf{40}(7), 2375--2379, (2008), doi:10.1016/j.physe.2007.07.007.

\bibitem{ando_environment_2010} Ando, Tsuneya, ``Environment {Effects} on {Excitons} in {Semiconducting} {Carbon} {Nanotubes}'', Journal of the Physical Society of Japan, \textbf{79}(2), 024706, (2010), doi:10.1143/JPSJ.79.024706.

\bibitem{le_louarn_aln_2009} Le Louarn, A., V{\'e}zian, S., Semond, F., Massies, J., ``{AlN} buffer layer growth for {GaN} epitaxy on (1 1 1) {Si}: {Al} or {N} first?'', Journal of Crystal Growth, \textbf{311}(12), 3278--3284, (2009), doi:10.1016/j.jcrysgro.2009.04.001.

\bibitem{park_effect_2014} Park, Young-Shin, Bae, Wan Ki, Padilha, Lazaro A., Pietryga, Jeffrey M., Klimov, Victor I., ``Effect of the {Core}/{Shell} {Interface} on {Auger} {Recombination} {Evaluated} by {Single}-{Quantum}-{Dot} {Spectroscopy}'', Nano Letters, \textbf{14}(2), 396--402, (2014), doi:10.1021/nl403289w.

\bibitem{schermelleh_super-resolution_2019} Schermelleh, Lothar, Ferrand, Alexia, Huser, Thomas, Eggeling, Christian, Sauer, Markus, Biehlmaier, Oliver, Drummen, Gregor P. C., ``Super-resolution microscopy demystified'', Nature Cell Biology, \textbf{21}(1), 72--84, (2019), doi:10.1038/s41556-018-0251-8.

\bibitem{iff_substrate_2017} Iff, Oliver, He, Yu-Ming, Lundt, Nils, Stoll, Sebastian, Baumann, Vasilij, H{\"o}fling, Sven, Schneider, Christian, ``Substrate engineering for high-quality emission of free and localized excitons from atomic monolayers in hybrid architectures'', Optica, \textbf{4}(6), 669--673, (2017), doi:10.1364/OPTICA.4.000669.
\end{thebibliography}
\end{document}